\documentclass[pra,aps,amsmath,amssymb,twocolumn,floatfix,superscriptaddress]{revtex4-2}
\usepackage{amsmath}
\usepackage{graphicx}
\usepackage{color}
\usepackage{amssymb}
\usepackage[utf8]{inputenc}
\usepackage[T1]{fontenc}

\usepackage{comment}
\usepackage[normalem]{ulem}
\usepackage{hyperref}
\hypersetup{colorlinks,
 linkcolor=blue,%
 citecolor=blue,%
 urlcolor=blue
}

\usepackage{array}

\begin{document}

\title{Double unstable avoided crossings and complex domain patterns formation in spin-orbit coupled spin-1 condensates}

\author{Sanu Kumar Gangwar}
\affiliation{Department of Physics, Indian Institute of Technology, Guwahati 781039, Assam, India} 

\author{Rajamanickam Ravisankar}
\affiliation{Department of Physics, Zhejiang Normal University, Jinhua 321004, PR China}
\affiliation{Zhejiang Institute of Photoelectronics \& Zhejiang Institute for Advanced Light Source, Zhejiang Normal University, Jinhua, Zhejiang 321004, China}

\author{Henrique Fabrelli}
\affiliation{Centro Brasileiro de Pesquisas Físicas, 22290-180 Rio de Janeiro, RJ, Brazil}

\author{Paulsamy Muruganandam}
\affiliation{Department of Physics, Bharathidasan University, Tiruchirappalli 620024, Tamil Nadu, India}
\affiliation{Department of Medical Physics, Bharathidasan University, Tiruchirappalli 620024, Tamil Nadu, India}

\author{Pankaj Kumar Mishra}
\affiliation{Department of Physics, Indian Institute of Technology, Guwahati 781039, Assam, India}

\begin{abstract}
We analyze the impact of spin-orbit and Rabi couplings on the dynamical stability of spin-orbit-coupled spin-1 Bose-Einstein condensates for ferromagnetic (FM) and antiferromagnetic (AFM) interactions. Determining the collective excitation spectrum through Bogoliubov-de-Gennes theory, we characterize the dynamical stability regime via modulational instability. For AFM interactions, the eigenspectrum reveals the presence of both stable and unstable avoided crossings (UAC), with the first-excited branch undergoing a double unstable avoided crossing. In contrast, with ferromagnetic interactions, only a single UAC, which occurs between the low-lying and first-excited branches, is observed. Furthermore, the eigenvectors demonstrate the transition from density-like to spin-like behaviour, as the collective excitation shows the transition from stable to unstable mode for both the FM and AFM interactions. In the multi-band instability state,  eigenvectors display spin-density mixed mode, while they show spin-flip nature in the avoided crossing regime. Our analysis suggests that spin-orbit coupling enhances the instability gain, while Rabi coupling plays the opposite role. Finally, we corroborate our analytical findings of stable and unstable regimes through numerical simulations of the dynamical evolution of the condensates by introducing the perturbations upon quenching the trap strength. The dynamical phases show the formation of complex domains with AFM interaction, which may be attributed to the double unstable avoided crossings in such a system.  

\end{abstract}

 \maketitle

\section{Introduction}
The experimental realization of spin-orbit-coupled (SOC) Bose-Einstein condensates (BECs) in the laboratory has opened up a wide range of phenomena to explore in ultracold atomic physics. Initially achieved in experiments using two of the three hyperfine components of the $F = 1$ states of ${}^{87}$Rb~\cite{Lin2011}, and later extended to spin-1 BECs~\cite{Campbell2016, Luo2016tun}, these systems have become a rich ground for exploring exotic quantum phenomena, such as superfluidity~\cite{Zhu2012}, supersolidity~\cite{Li2017}, metastable supersolid~\cite{Xia2023}, modulation instability~\cite{Bhat2015, Bhuvaneswari2016, Limod2017}, vortices~\cite{Radi2011, Adhikari2020}, solitons~\cite{Achilleos2013, Xu2013}, etc.
 
Numerical simulations have played a crucial role in revealing many fascinating properties of spin-orbit-coupled Bose-Einstein condensates (SOC-BECs), primarily through the use of the mean-field Gross-Pitaevskii (GP) equation~\cite{Ohmi1998, Ho1998, Wang2007, Bao2008, Lim2008, Muruganandam2009, Ravisankar2021}. Liu {\it et al.} reported exact soliton solutions and manifold mixing dynamics in quasi-one-dimensional SOC spin-1 BECs~\cite{Liu2014}. Numerical studies along similar lines suggest the emergence of more complex phases, such as multiring structures, stripes, and superlattice solitons, in quasi-two-dimensional SOC spin-1 BECs~\cite{Adhikari2021}. In addition to these phases, the formation of symbiotic spinor solitons has been reported in quasi-one-dimensional and quasi-two-dimensional spin-1 ferromagnetic (FM) BECs~\cite{Adhikariss2021}. For antiferromagnetic (AFM) SOC spin-1 BECs, the formation of stable multi-peak vector solitons in quasi-one-dimensional systems has been reported~\cite{Adhikari2020stable}. Recently, He {\it et al.}, through linear stability analysis, demonstrated the existence of stationary and moving bright solitons in quasi-one-dimensional SOC spin-1 BECs under the influence of a Zeeman field~\cite{He2023}.


Collective excitations, which are low-lying excitations of the quantum gas, are key in determining the stability properties of ground-state phases and the behaviour of fluctuations and superfluidity of the BECs. Landau developed a framework for studying elementary excitations in superfluid helium, while for weakly interacting Bose gases, Bogoliubov derived the theory of elementary excitations~\cite{Bogoliubov1947}. Numerous theoretical works have reported the collective excitations in single-component BECs~\cite{Fetter1996, Edwards1996, Singh1996, Stringari1996}. Goldstein {\it et al.}~\cite{Goldstein1997} extended the Hartree-Bogoliubov theory for multi-component BECs and derived the quasi-particle frequency spectrum. They further demonstrated that interferences arising from cross-coupling between condensate components resulted in a reversal of the sign of the effective two-body interaction and the onset of spatial instabilities. The stability of supercurrents in BECs with one-dimensional SOC has also been studied, showing that supercurrents in the plane-wave phase exhibit dynamical instability. Additionally, extensive energetic instability analysis of supercurrent states has been reported~\cite{Ozawa2013}. Several laboratory experiments have reported collective excitations, including phonon-like excitations in a dilute gas of ${}^{87}$Rb~\cite{Jin1996}, and collective excitations of sodium atoms in a magnetic trap~\cite{Mewes1996}. Khamehchi {\it et al.}~\cite{Khamehchi2014} used Bragg spectroscopy to measure the collective excitations in SOC-BECs, revealing the presence of phonon-maxon-roton modes, as predicted using the Bogoliubov-de Gennes (BdG) theory. In a recent study, Ravisankar {\it et al.}~\cite{Mishra2021} investigated the influence of spin-orbit (SO) and Rabi coupling strengths on the dynamical instability of quasi-two-dimensional binary BECs using Bogoliubov theory. Their analysis revealed the presence of phonon, roton, and maxon modes, demonstrating that increasing the SO coupling enhances instability, whereas stronger Rabi coupling stabilizes the system.

Modulation instability (MI) is a measure of instability in BECs, which becomes one of the key criteria for determining the stability of the condensate. In recent years, numerous studies have been carried out to examine the stability of the spinor condensate using MI. For instance, Tsubota {\it et al.}~\cite{Tsubota2004, Tsubota2006} demonstrated that modulational instability (MI) in two-component BECs leads to the formation of multiple domains and regions dominated by the solitary waves. Robins et al.~\cite{Robins2001} and Zhang et al.~\cite{Zhang2005} performed numerical MI analyses on the FM and AFM phases of spin-1 BECs, revealing that the FM phase is unstable, while the AFM phase is stable. Some works show the stability analysis of these phases in the presence of the external field. For instance, Matuszewski { \it et al.} showed that a homogeneous magnetic field induces spatial MI in AFM spin-1 BECs, resulting in the formation of spin domains in sodium condensates confined in optical traps.~\cite{Matus2008}. In a subsequent study, they found that the metastable phases of an antiferromagnetic spin-1 condensate, in a simple model with pure contact interactions, could exhibit a roton-like minimum in the excitation spectrum. The presence of an external magnetic field gives rise to the instability of the roton modes, which can lead to the spontaneous emergence of regular periodic patterns~\cite{Matuszewski2010}. Similarly, Kronj\"ager \textit{et al.} reported spontaneous pattern formation in antiferromagnetic spinor BECs and identified several linearly unstable modes using a mean-field approach~\cite{Kronjager2010}.
In other direction, Pu \textit{et al.} reported that magnetic-field-induced dynamical instabilities in spin-1 BECs are accompanied by $I_o$ type instabilities in the presence of nonzero magnetic field~\cite{Pu2016}. Systems with such embedded instabilities spontaneously develop a spatial pattern on time evolution from the uniform initial state. 



On the other hand, studies on instabilities in SOC-BECs remain relatively limited. Bhuvaneswari {\it et al.}~\cite{Bhuvaneswari2016} theoretically investigated modulational instability (MI) in quasi-two-dimensional SOC-BECs with Rabi coupling, assuming equal densities for both pseudo-spin components. They found that unstable modulations arise from initially miscible condensates, depending on the combination of signs of intra- and intercomponent interaction strengths. The SOC enhances instability regardless of the interaction type; however, in the case of attractive interactions, SOC further amplifies the MI. In quasi-one-dimensional systems, Bhat {\it et al.}~\cite{Bhat2015} demonstrated that two-component SOC-BECs with equal component densities exhibit MI, leading to the formation of a striped phase as the ground state. Additionally, {\it Li et al.}~\cite{Limod2017} studied MI in quasi-one-dimensional SOC-BECs with Raman coupling, showing that instability occurs for repulsive density-density and spin-exchange interactions even in the absence of SOC and Raman coupling.
 

Identifying unstable phases and their underlying mechanisms in complex spin-1 BECs remains a significant challenge. For instance, a recent study by Gangwar {\it et al.} reported the occurrence of unstable avoided crossings (UACs) in the context of FM interactions in spin-1 SOC-BECs~\cite{Gangwar2024}. However, a comprehensive understanding of the interplay among interaction strength, SOC, and Rabi coupling, the key factors in the emergence of complex phases, remains elusive. In this paper, we present a detailed analysis of modulational instabilities in SOC-coupled spin-1 BECs, exploring how interaction types (AFM and FM) and coupling parameters influence the eigenspectrum and resulting dynamics. For AFM interactions, the eigenspectrum exhibits both stable and unstable avoided crossings (UACs) between the low-lying and first-excited branches, as well as between the first-excited and second-excited branches. Notably, the first-excited branch under AFM interactions displays a double UAC, a phenomenon previously reported for spin-1 BECs in the presence of a magnetic field~\cite{Pu2016, Kronjager2010}. In this study, however, we explicitly demonstrate that this double unstable avoided crossing is induced by SOC. For FM interactions, we find that a single UAC occurs between the low-lying and first-excited branches, with a double UAC appearing only when the Rabi coupling strength is negative. The UAC is associated with the \(I_o\)-type instability, which drives pattern formation in the density profile of the condensate~\cite{Bernier2014}. To complement our analytical findings, we conduct numerical simulations for both AFM and FM interactions. These simulations provide deeper insight into the dynamics and pattern formation arising from modulational instability.

The structure of our paper is as follows. In section~\ref{sec:2}, we present the mean-field model to explore MI of SOC spin-1 BECs with Rabi-coupling. Following this, we provide a detailed theoretical formalism using BdG to calculate the collective excitation spectrum in Section~\ref{sec:3}. In Sec.~\ref{sec:4}, we present the collective excitation spectrum for FM interactions, followed by an analysis for AFM interactions in Sec.~\ref{sec:5}. In Sec.~\ref{sec:6}, we present the effect of interaction strengths on MI. In Sec.~\ref{sec:7}, we demonstrate the variation of the Band gaps between the low-lying and first excited states and between the first and second excited states for FM and AFM interaction. In  Sec.~\ref{sec:8}, we present numerical simulations using the GPEs, which complement the dynamical instability regions obtained from the BdG analysis. Finally, in Section~\ref{sec:9}, we summarize our findings.
\section{Mean-field model}
\label{sec:2}
We consider a quasi-one dimensional  SO coupled spin-1 BECs with tight confinement in the transverse direction~\cite{Salas2002}, which can be described using three sets of coupled GPEs, given as~\cite{Gautam2015, Katsimiga2021, Adhikariss2021, Ueda2012},
\begin{widetext}
\begin{subequations}
\begin{align}%
\mathrm{i} \frac{\partial \psi_{\pm1}}{\partial t} =&  \bigg[- \frac{1}{2 } \frac{\partial^{2}}{\partial x^{2}}+ V (x)+c_{0}\rho\bigg] \psi_{\pm1} \mp \frac{ k_{L}}{\sqrt{2}} \frac{\partial \psi_{0}}{\partial x} + c_{2}\bigg(\rho_{\pm1} + \rho_{0} - \rho_{\mp1}\bigg) \psi_{\pm1} + c_{2} \psi_{0}^{2} \psi_{\mp1}^{*}+\frac{\Omega}{\sqrt{2}}\psi_{0}, \label{gpe1} \\ 
\mathrm{i} \frac{\partial \psi_{0}}{\partial t} =&  \bigg[- \frac{1}{2} \frac{\partial^{2}}{\partial x^{2}}+ V (x) +c_{0}\rho\bigg] \psi_{0} + \frac{ k_{L}}{\sqrt{2}} \bigg[ \frac{\partial \psi_{+1}}{\partial x}  - \frac{\partial \psi_{-1}}{\partial x}\bigg]   + c_{2}( \rho_{+1} + \rho_{-1})\psi_{0} + 2 c_{2} \psi_{0}^{*} \psi_{+1} \psi_{-1} +\frac{\Omega}{\sqrt{2}}(\psi_{1}+\psi_{-1}) \label{gpe2}
\end{align}%
\end{subequations}
\end{widetext}
Here, $\psi_{+1}$, $\psi_{0}$, and $\psi_{-1}$ are the spinor wavefunctions that satisfy the normalization condition $\int_{-\infty}^{\infty} d x\; \rho  = 1$ with $\rho = \sum_{j=-1}^{1} \rho_{j}$, total atomic density of the condensate, and $\rho_{j}$ = $\vert\psi_{j}(x) \vert^2$ represents the density of $j$-th component of the condensate with j = $\pm1, 0$. The Eqs. (\ref{gpe1}) and (\ref{gpe2}) are non-dimensionalized, using time, length, and energy $t = \omega_{x} \tilde{t}$, $x = \tilde{x}/l_{0}$, and $\hbar \omega_{x}$, respectively. The resulting condensate wavefunction is $\psi_{\pm 1,0} = \sqrt{\frac{l_{0}}{N}} \tilde{\psi}_{\pm 1,0}$, where, $l_{0} = \sqrt{\hbar / m \omega_{x}}$, is the oscillator length for the trap frequency $\omega_{x}$ along the x-axis. The trap strength is given by $V (x) = x^{2} /2$, density-density interaction strength $c_{0} = 2 N l_{0}(a_{0} + 2a_{2})/ 3 l_{\perp}^{2}$, spin-exchange interaction strength $c_{2} = 2 N l_{0}(a_{2} - a_{0})/ 3 l_{\perp}^{2}$, where $a_{0}$ and $a_{2}$ are the s-wave scattering lengths in total spin channels $0$ and $2$, respectively. Upon tuning $c_{2} < 0$ ($ c_2> 0$), one can have the FM (AFM) interaction of the condensates~\cite{Ueda2012, Stamper2013}. Here, $l_{\perp} = \sqrt{ \hbar / (m \omega_{\perp})}$ is the oscillator length in transverse direction with $\omega_{\perp} = \sqrt{\omega_{y}\omega_{z}}$. The SO and Rabi coupling strengths are given by  $k_{L} = \tilde{k_{L}} / \omega_{x} l_{0}$, $\Omega = \tilde{\Omega} / (\hbar \omega_{x})$, respectively. In the above description, the quantities with the tilde represent dimensionful quantities. In this entire work, we consider all parameters to be dimensionless. 
The energy functional corresponding to the coupled GP equations (\ref{gpe1})-(\ref{gpe2}) is given by~\cite{Ravisankar2021},
\begin{align}
    E  = & \frac{1}{2} \int dx \bigg\{ \sum_{j} \left\vert \partial_{x} \psi_{j} \right\vert^{2} + 2 V(x) \rho + c_{0} \rho^{2}  \nonumber  \\  & + c_{2}[ \rho_{+1}^{2} + \rho_{-1}^{2} + 2( \rho_{+1}\rho_{0}+\rho_{-1}\rho_{0} -\rho_{+1}\rho_{-1} \nonumber  \\  & +\psi_{-1}^{*}\psi_{0}^{2}\psi_{+1}^{*}+ \psi_{-1}\psi_{0}^{*2}\psi_{+1})]  + \sqrt{2} \Omega[(\psi_{+1}^{*}+\psi_{-1}^{*})\psi_{0} \nonumber  \\  & +\psi_{0}^{*}(\psi_{+1}+\psi_{-1})]+ \sqrt{2} k_{L}[  (\psi_{-1}^{*}-\psi_{+1}^{*})  \partial_{x} \psi_{0}  \nonumber  \\  & + \psi_{0}^{*}(\partial_{x} \psi_{+1}- \partial_{x}\psi_{-1})]\bigg\} \label{eqn5}
\end{align}

Here, we outline the experimentally viable range for the parameters that we have considered for our simulations. For the FM interaction, we consider $^{87}$Rb BECs with $N \sim 2\times 10^{4}$ number of atoms. The axial trap frequency as $\omega_x = 2 \pi \times 50$ Hz and the transverse trap frequencies as $\omega_y = \omega_z = 2 \pi \times 500$ Hz. The resultant characteristic lengths would be $l_{0} = 1.52 \mu m$, and $l_{\perp} = 0.48 \mu m$. For the AFM interaction, we consider the BECs of $^{23}$Na atoms. The resultant characteristic lengths for $^{23}$Na BECs are $l_{0} = 2.9 \mu m$, and $l_{\perp} = 0.9 \mu m$. The spin-dependent and spin-independent interactions can be achieved by controlling the $s$-wave scattering lengths through Feshbach resonance~\cite{Inouye1998, Marte2002, Chin2010}. However, the SOC strength $k_L = \{ 0.1 - 5 \}$ can be tuned by changing the laser wavelengths in the range of $\{68.86 \mu$m - $1377.22 \mbox{nm}\}$. However, the dimensionless Rabi frequency interval $\Omega=[0,5]$ used in the simulation can be attained by tunning the Raman laser strength in the range of  $2\pi \hbar\times \{5 - 250\}$ Hz.
\section{Collective Excitation spectrum}
\label{sec:3}
In this section, we present the collective excitation spectrum of SOC spin-1 BECs using the Bogoliubov-de-Gennes (BdG) analysis. We consider the uniform ground state wavefunction ($\phi_{j}$) is perturbed by the wavefunction $\delta \phi_{j}$ for which the excited state wavefunction is given by~\cite{Pu2016, Zhu2020},
\begin{align}\label{excwave}
\psi_{j}(x,t) = e^{-\mathrm{i} \mu_{j} t}[\phi_{j} +  \delta \phi_{j}(x, t) ]
\end{align}    
where,
\begin{align}\label{pertwave}
\delta \phi_{j}(x,t) = u_{j} e^{\mathrm{i}( q_{x} x - \omega t)} + v_{j}^{*} e^{- \mathrm{i}(q_{x} x - \omega^{*} t)}
\end{align}%
and the uniform ground state wavefunction is considered  as $\phi_{j} = (1/2, -1/\sqrt{2},  1/2)^{T}$, $\mu_{j}$ denotes the chemical potential, and $u_{j}$,  $v_{j}$ are the Bogoliubov amplitudes corresponding to the three spin-components. As we substitute Eq.~(\ref{excwave}) in the dynamical Eqs.~(\ref{gpe1})-(\ref{gpe2}), we obtain
\begin{align}\label{bdgeigprbm}
(\mathcal{L} -\omega \mathrm{I}) 
\begin{pmatrix}
    u_{+1} & v_{+1} & u_{0} & v_{0} & u_{-1} & v_{-1}
\end{pmatrix}^{T} = 0,
\end{align}
where $T$ represents the transpose of the matrix, $\mathrm{I}$ is a $6 \times 6$ identity matrix and  $\mathcal{L}$ is $6 \times 6$ matrix given by, 
\begin{widetext}
\begin{align} 
\label{bdgmatrix}
\mathcal{L} =
 \begin{pmatrix}
 H_{+}-\mu_{+} & \mathcal{L}_{12} & \mathcal{L}_{13}
 & \mathcal{L}_{14} & \mathcal{L}_{15} &
 \mathcal{L}_{16}\\
 \mathcal{L}_{21} & -H_{+}+\mu_{+} & \mathcal{L}_{23} & \mathcal{L}_{24}
& \mathcal{L}_{25} & \mathcal{L}_{26}\\
 \mathcal{L}_{31} & \mathcal{L}_{32} & H_{0}-\mu_{0} & \mathcal{L}_{34} & \mathcal{L}_{35} & \mathcal{L}_{36}\\
 \mathcal{L}_{41} & \mathcal{L}_{42} & \mathcal{L}_{43} & -H_{0}+\mu_{0} & \mathcal{L}_{45} & \mathcal{L}_{46}\\
 \mathcal{L}_{51} & \mathcal{L}_{52} & \mathcal{L}_{53} & \mathcal{L}_{54} & H_{-} -\mu_{-} & \mathcal{L}_{56}\\
 \mathcal{L}_{61} & \mathcal{L}_{62} & \mathcal{L}_{63} & \mathcal{L}_{64} & \mathcal{L}_{65} & -H_{-}+\mu_{-}
 \end{pmatrix}
 \end{align}
 \end{widetext}
The matrix elements of $\mathcal{L}$ have been provided in the Appendix.~\ref{matrx:BdG}. Bogoliubov coefficients follow the normalization condition,
\begin{align}\label{normbdg}
 \int \bigg(\sum_{j} \{\left\vert u_{j} \right\vert^{2} - \left\vert v_{j} \right\vert^{2}\}\bigg) dx = 1
\end{align}
We compute the determinant of the matrix $\mathcal{L}$ and equate it to zero $det (\mathcal{L} - \rm{I} \omega ) = 0$, which yields the characteristic equation given as,   
\begin{align}\label{bdgex}
\omega^{6}+b \omega^{4}+ c \omega^{2} + d =0
\end{align}
where the coefficients $b$, $c$, and $d$ are supplied in the Appendix.~\ref{matrx:BdG}.%

By numerically solving the BdG equations~(\ref{bdgmatrix}), we complement the reliability of the analytical results for the excitation spectrum and obtain the eigenvectors as a function of $q_x$. For the numerical calculation, we have considered the real space $[-1000:1000]$ grid with a step size of $h_x=0.2$. We employ the Fourier collocation method, which utilizes the LAPACK package~\cite{Anderson1999} to diagonalize the truncated BdG matrix we acquire by numerically computing the Fourier transform of the BdG equations. In momentum space $q_{x}$, we consider $[-700:700]$ modes, with a grid step size of $h_{q_x}=0.0157$.

In our recent work, we conducted a detailed analysis of the collective excitation spectrum for FM interactions in SOC spin-1 BECs. We reported the presence of unstable avoided crossings between the low-lying and first excited states of the eigenspectrum~\cite{Gangwar2024}. In this paper, we extend this analysis to antiferromagnetic (AFM) interactions, where we observe evidence of double avoided crossings. In the following section, first, we discuss the effects of SOC and Rabi coupling on the eigenspectrum for FM interactions and subsequently, we extend the analysis to AFM interactions.
\subsection{Collective excitation spectrum of FM SOC-BECs ($c_{0} > 0$, $c_{2} < 0$)}
\label{sec:4}
\begin{figure*}[!htp]
\begin{centering}
\centering\includegraphics[width=0.99\linewidth]{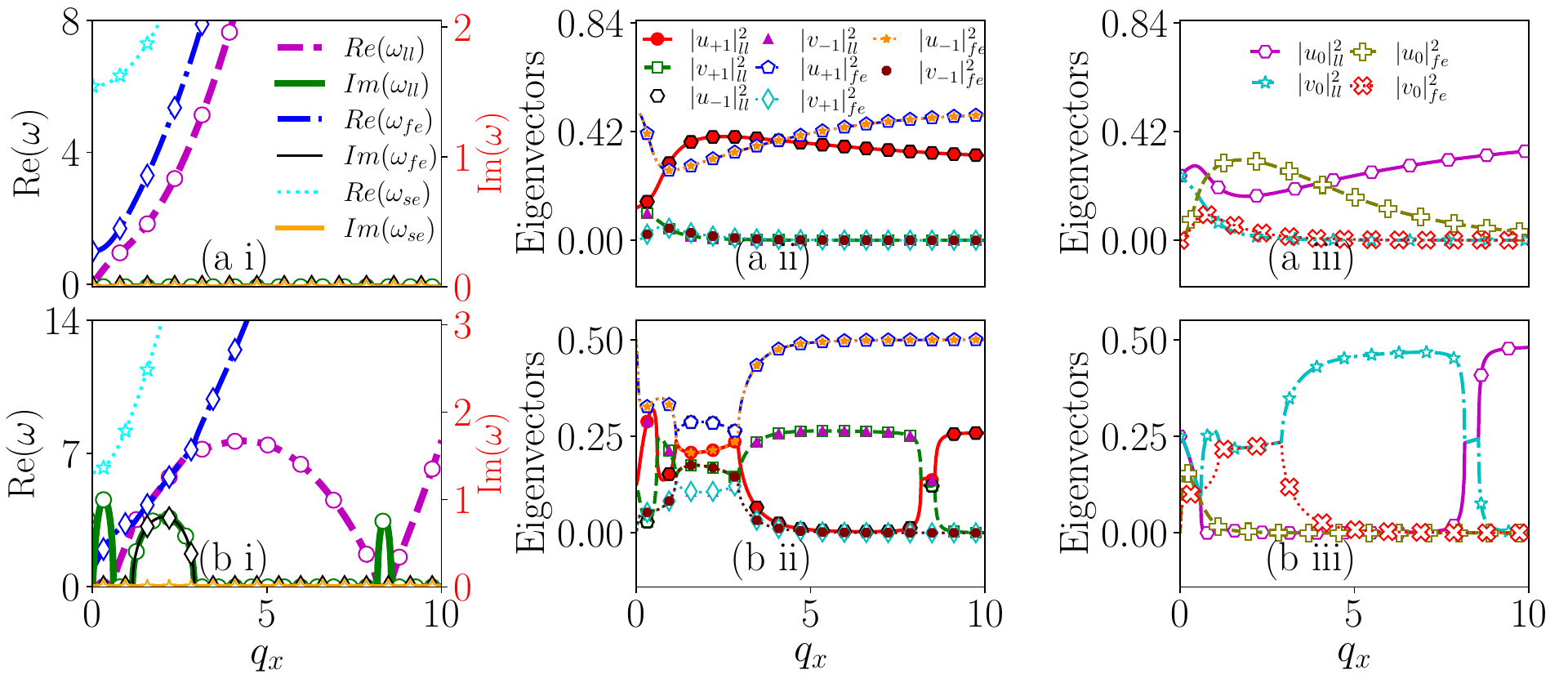}
\caption{Eigenvalue and corresponding eigenvectors  for (a) ($k_{L}, \Omega$) = ($0.5, 1.0$) showing real eigenfrequencies and (b) ($k_{L}, \Omega$) =($4.5, 1.0$) showing the multiband instability for FM interaction $c_{0} = 5$, $c_{2} = -2.0$. The dashed magenta line represents $\text{Re}(\omega_{ll})$, the thick green line represents $\vert  \text{Im}(\omega_{ll})\vert$, dash-dotted blue line represents $\text{Re}(\omega_{fe})$, the solid black line represents $\vert \text{Im}(\omega_{fe})\vert$, dotted cyan line represents $\text{Re}(\omega_{se})$, and the solid orange line represents $\vert \text{Im}(\omega_{se})\vert$, here solid and dash-dotted lines are the analytical results of BdG Eq.~\ref{bdgex} and open circles are numerical results obtained by solving Eq.~\ref{bdgeigprbm}. 
Eigenvectors of low-lying branch are $\vert u_{+1}\vert^{2}_{ll}$ (red dot), $\vert u_{-1}\vert^{2}_{ll}$ (black hexagons), $\vert u_{0}\vert^{2}_{ll}$ (magenta hexagons), $\vert v_{0}\vert^{2}_{ll}$ (cyan open stars), $\vert v_{+1}\vert^{2}_{ll}$ (green open squares), and $\vert v_{-1}\vert^{2}_{ll}$ (magenta triangles). For the first-excited branch, $\vert u_{+1}\vert^{2}_{fe}$ (blue open pentagon), $\vert u_{-1}\vert^{2}_{fe}$ (orange stars), $\vert u_{0}\vert^{2}_{fe}$ (olive open plus), $\vert v_{0}\vert^{2}_{fe}$ (red open x with the dotted line), $\vert v_{+1}\vert^{2}_{fe}$ (cyan open diamond), and $\vert v_{-1}\vert^{2}_{fe}$ (maroon dots). (a i) exhibits only real eigenfrequencies, and (b i) depicts multi-band instability, which is symmetric in the quasi-momentum direction. Corresponding eigenvectors depict density-like mode and spin-like mode, respectively.  The UAC among branches results in $I_{o}$ type instability band in the eigenspectrum. The vertical scale on the right represents the magnitude of imaginary eigenfrequencies. 
}   
\label{t-fig0}
\end{centering}
\end{figure*}

In this section, we present the collective spectrum analysis for SOC Bose-Einstein condensates (BECs) with ferromagnetic (FM) interaction, characterized by $c_{0} = 5.0$ and $c_{2} = -2.0$. The negative ($\omega_{-}$), positive ($\omega_{+}$), and zeroth ($\omega_{0}$) branches are designated as the low-lying ($\omega_{ll}$), first-excited ($\omega_{fe}$), and second-excited ($\omega_{se}$) branches of the eigenspectrum, respectively. It has been shown that the eigenspectrum of SOC BECs exhibits intriguing features, including $I_{o}$-type instabilities~\cite{Bernier2014, Pu2016}. Generally, these $I_{o}$-type instabilities arise from unstable avoided crossings (UACs) between the branches.


We begin our analysis by first calculating the eigenvalues of the BdG matrix, as given in Eq.~\ref{bdgmatrix}. In Fig.~\ref{t-fig0}, we present the eigenspectrum corresponding to regime I ($k_{L}^{2} < \Omega$) (top panel) and regime II ($k_{L}^{2} > \Omega$) (bottom panel) of the condensate. In regime I ($k_{L}^{2} < \Omega$), we observe only real eigenfrequencies, accompanied by a gap between all branches. The low-lying branch exhibits a phonon mode, as shown in Fig.~\ref{t-fig0}(a i). The eigenspectrum is symmetric about the quasi-momentum direction. To further explore the detailed nature of the excitations, we plot the eigenvectors corresponding to the eigenspectrum in Fig.~\ref{t-fig0}(a i) in Figs.~\ref{t-fig0}(a ii) and \ref{t-fig0}(a iii). Here, we display three sets of eigenvector components, namely, $\vert u_{+1} \vert^{2}$, $\vert v_{+1} \vert^{2}$, $\vert u_{0} \vert^{2}$, $\vert v_{0} \vert^{2}$, $\vert u_{-1} \vert^{2}$, and $\vert v_{-1} \vert^{2}$, for a specific branch of the eigenspectrum. We find that all eigenvector components exhibit a density-like mode consistent with the real eigenfrequencies observed, aligning with previous studies on spin-1/2 SOC BECs~\cite{Abad2013, Mishra2021}. Additionally, these eigenvectors satisfy the following relations for the low-lying and first-excited branches:
\begin{align} \label{eq:density}
& \vert u_{+1} \vert_{ll}^{2} - \vert u_{-1} \vert_{ll}^{2} = 0, \quad
\vert v_{+1} \vert_{ll}^{2} - \vert v_{-1} \vert_{ll}^{2} = 0, \notag \\    
& \vert u_{+1} \vert_{fe}^{2} - \vert u_{-1} \vert_{fe}^{2} = 0, \quad
\vert v_{+1} \vert_{fe}^{2} - \vert v_{-1} \vert_{fe}^{2} = 0.
\end{align}
Apart from this, we find that the eigenvector components of the low-lying branch appear to approach the same value at $q_x \approx 0$, a typical feature of the phonon mode in the eigenspectrum [see Figs.~\ref{t-fig0}(a ii, a iii)]. In Fig.~\ref{t-fig0}(b i), we present the eigenfrequency for regime II ($k_L^2 > \Omega$) of the condensate, which clearly reveals the presence of imaginary eigenfrequencies in the low-lying and first-excited branches of the eigenspectrum, whereas the second-excited branch exhibits only real eigenfrequencies. The low-lying branch displays three instability bands at wave number and corresponding eigenfrequency pairs given by $\{q_x, \omega\} = \{0.33, 0.999\}$, $\{2.04, 0.799\}$, and $\{8.37, 0.773\}$, while the first-excited branch exhibits a single instability band at $q_x = 2.04$ with an eigenfrequency of $\omega = 0.799$. This single-band instability in the first-excited branch arises due to an UAC between the low-lying and first-excited branches of the eigenspectrum, occurring within the quasi-momentum range $1.14 \lesssim q_x \lesssim 2.94$. This interaction is primarily responsible for the $I_o$-type dynamical instability~\cite{Bernier2014, Pu2016, Gangwar2024}. The presence of imaginary eigenfrequencies indicates dynamical instability in the condensate phase. Similar to regime I, in regime II, we also observe that the eigenspectrum is symmetric about the quasi-momentum direction; however, it distinctly exhibits multi-band instability. This multi-band nature manifests in the eigenvectors as a spin-like mode, satisfying a specific relationship between the low-lying and first-excited branches of the excitation spectrum.
\begin{align}\label{eq:spin}
& \vert u_{+1} \vert_{ll}^{2} - \vert u_{-1} \vert_{ll}^{2} \neq 0, \vert v_{+1} \vert_{ll}^{2} - \vert v_{-1} \vert_{ll}^{2} \neq 0,  \notag \\ 
& \vert u_{+1} \vert_{fe}^{2} - \vert u_{-1} \vert_{fe}^{2} \neq 0, \vert v_{+1} \vert_{fe}^{2} - \vert v_{-1} \vert_{fe}^{2} \neq 0.     
\end{align}

The eigenvectors of the low-lying branch display an interesting feature of the transition from spin-like mode to density-like mode which indicates the presence of transition from $\text{Im}(\omega) \to \text{Re}(\omega)$ and then density-like mode to spin-like mode informs that $\text{Re}(\omega) \to \text{Im}(\omega)$ and in the weak wavelength limit only $\text{Re}(\omega)$ is present and $\text{Im}(\omega)$ is absent. This particular feature suggests the presence of spin-density-spin mixed mode of the eigenvector in the three instability bands regions of the eigenspectrum. The first-excited branch of the eigenspectrum has a single instability band that occurs due to the UAC between the low-lying and first-excited branches, which lies in the quasimomentum range $1.14 \lesssim q_{x} \lesssim 2.94$. Closer to the $q_x$ of UAC, the eigenvectors of the low-lying and first-excited branches display out-of-phase behaviour among the branches [see Fig.~\ref{t-fig0}(b ii)]. However, the zeroth component of eigenvectors exhibits density-like mode only [see Fig.~\ref{t-fig0}(b iii)]. 


\begin{figure}
\begin{centering}
\centering\includegraphics[width=0.99\linewidth]{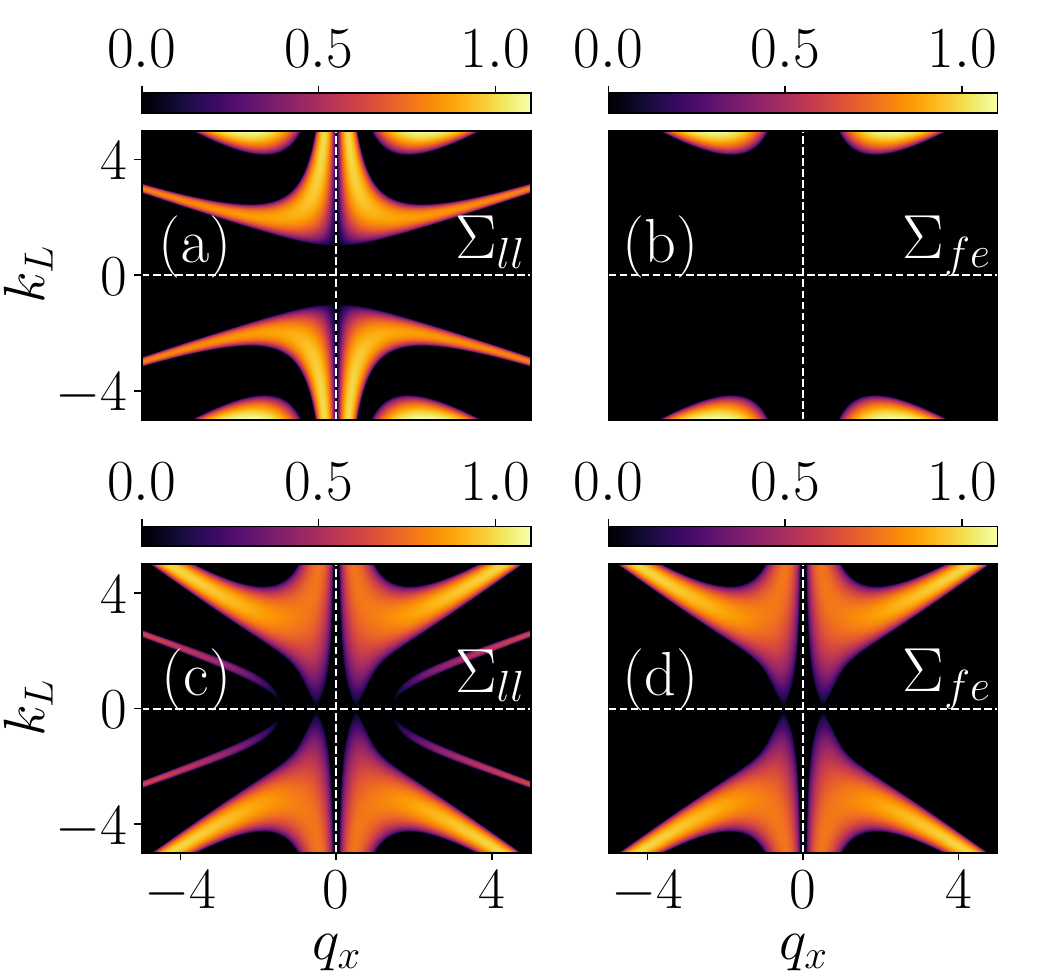}
\caption{Pseudo color plot of modulational instability gain for the ferromagnetic phase ($c_{0} = 5$, $c_2 = -2$) in $k_{L} - q_{x}$ plane for (a, b) $\Omega = 1$ and  (c, d) $\Omega = -1$. Left column panel corresponds to $\mathrm{\Sigma}_{ll} \equiv \mathrm{\vert Im(\omega}_{ll}) \vert$ and right column panel corresponds to $\mathrm{\Sigma}_{fe} \equiv \mathrm{\vert Im(\omega}_{fe}) \vert$. In (a), multi-band instability gain appears at $k_{L} = 1$, and in (b) a single-band instability appears at $k_{L} \approx 4.0$. In (c) and (d) double and single band instabilities appear at $k_{L} > 0.25$, respectively. The second-excited branch of the eigenspectrum shows the absence of instability gain for both sets of parameters. The instability gain is symmetric about $k_{L}$ and $q_{x}$.}
 \label{t-fig2}
\end{centering}
\end{figure}%



After analyzing the collective excitation spectrum, we wish to understand the appearance of MI in the phase plots. Specifically, we focus on the $k_L - q_x$ plane for the FM SOC-BECs. In Figs.~\ref{t-fig2}(a, b), we present the pseudo color representation of the modulational instability (MI) gain, defined as imaginary part of the modulus of eigenfrequency, for a Rabi coupling strength $\Omega = 1$ and with FM interaction strengths $c_{0} = 5.0, c_{2} = -2.0$. We observe that instability bands start emerging in the range where $k_{L}^2 > \Omega$, first appearing in the low-lying branch ($\Sigma_{ll}$) as a single band. However for $k_{L} > 3$ two instability band emerge and for $k_{L} > 4$ three bands are observed[see fig.~\ref{t-fig2}(a)]. 
For the first-excited branch ($\Sigma_{fe}$), instability emerges only at higher values of the SOC ($k_L \gtrsim 4.0$). This can be attributed to the emergence of a UAC between the $\omega_{ll}$ and $\omega_{fe}$ branches within the quasi-momentum range $0.95 < q_x < 3.64$. Initially, this forms a single instability band, which expands as the SOC strength increases. In Figs.~\ref{t-fig2}(c) and \ref{t-fig2}(d), we present the MI gain for the low-lying and first-excited modes, respectively, at $\Omega = -1$, with all other parameters held constant as those used for $\Omega = 1$. Here, we observe a markedly different trend compared to those results obtained with a positive Rabi frequency. In a spin-1 condensate, negative Rabi coupling increases the system’s energy, manifesting as a higher number of instability bands~\cite{Ravisankar2020}. For negative Rabi strength, we notice that the instability appears at a lower SOC value ($k_L > 0.25$) compared to positive Rabi strength [see Figs.~\ref{t-fig2}(c) and \ref{t-fig2}(d)]. The instability manifests as double bands in the low-lying branch and a single band in the first-excited branch, both of which intensify with increasing SOC strength. The instability gain is symmetric about the quasi-momentum $q_{x}$ and the SO coupling strength $k_{L}$. Notably, the second-excited branch exhibits no instability gain in the FM BECs, even in the presence of SOC ($k_L$).
\begin{figure}
\begin{centering}
\centering\includegraphics[width=0.99\linewidth]{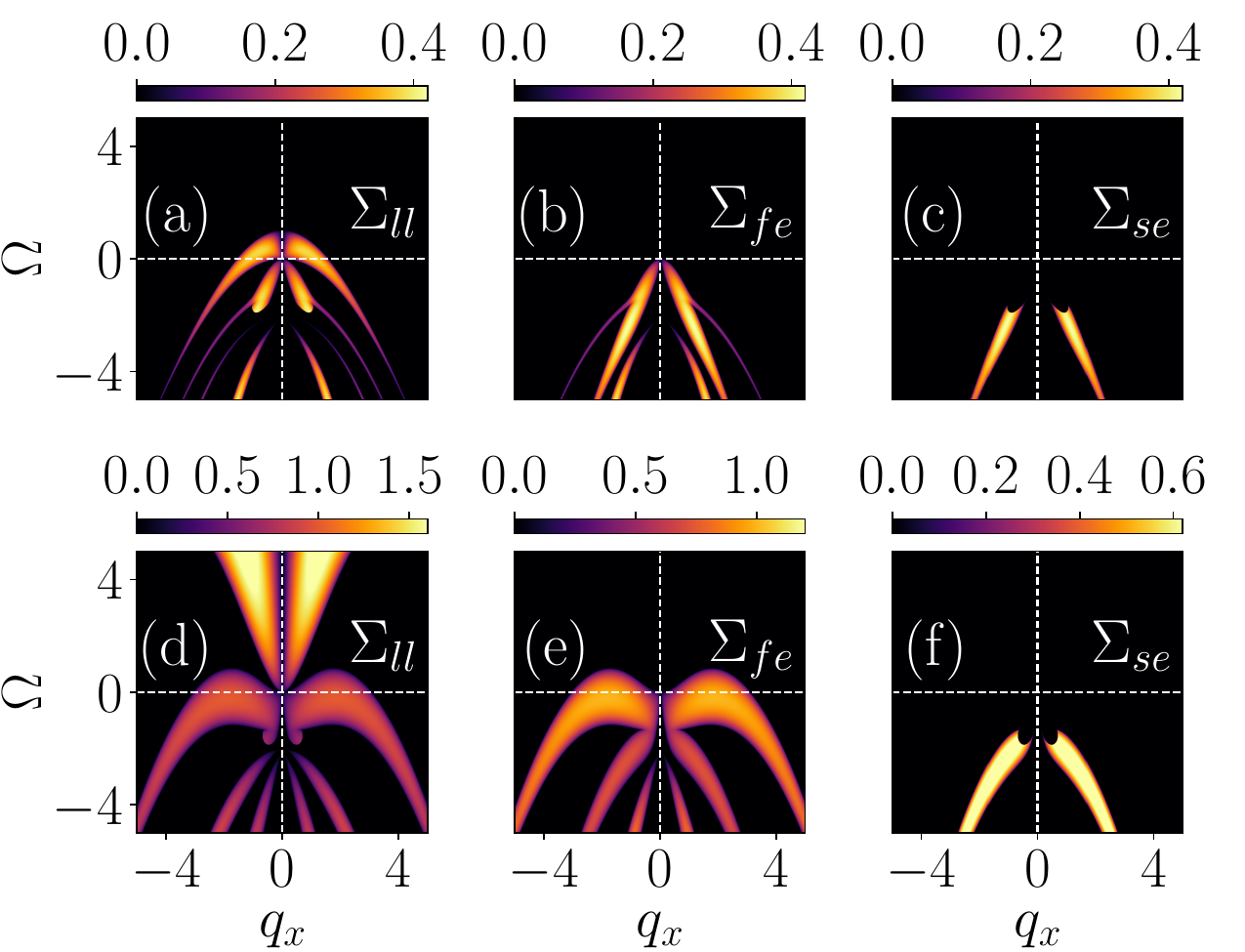}
\caption{{Pseudo color plot of the Modulational instability gain in $\Omega - q_{x}$ phase plane for FM interactions ($c_{0} = 5.0$, and $c_2 = -2.0$) with (a-c) for $k_{L} = 1$, and (d-f) for $k_{L} = 4.0$. (a,d) represents $\mathrm{\Sigma}_{ll} \equiv \mathrm{\vert Im(\omega}_{ll}) \vert$, (b,e) represents $\mathrm{\Sigma}_{fe} \equiv \mathrm{\vert Im(\omega}_{fe}) \vert$ and (e,f) for $\mathrm{\Sigma}_{se} \equiv \mathrm{\vert Im(\omega}_{se}) \vert$  low-lying, first-excited, and second-excited branches of the excitation spectrum. In (a), the primary band appears for $\Omega > 0$ following the relation $k_{L}^{2} = \Omega$, while in (b) and (c), it appears only for $\Omega < 0$, with a cutoff in (c), $\Omega \approx -1.66$. Similar to the top row, the primary instability band in (d) appears for $\Omega > 0$. In (d) and (f), the secondary and primary bands appear for $\Omega > 0$ different than the top row. In (f), single band instability appears for $\Omega < 0$, following a cutoff value $\Omega \approx -1.44$. The instability gain appears is not symmetric about $\Omega$ and increases upon decreasing the $\Omega$. However, it shows symmetric behaviour about $q_{x}$.}}
 \label{t-fig4}
\end{centering}
\end{figure}%
After getting a clear understanding of how Rabi coupling affects the MI gain in the $k_L -q_x$ plane, we now shift our focus on exploring the same in the $\Omega - q_x$ plane by fixing the SOC strength as $k_L = 1$.
We analyze the impact of Rabi coupling on the MI gain for the FM interaction strengths $c_{0} = 5.0$, $c_{2} = -2.0$. To systematically investigate the impact of Rabi coupling strength on the eigenspectrum, we begin our analysis by choosing $k_{L} = 1.0$. We find that the low-lying branch of the eigenspectrum exhibits finite instability gain for $\Omega < 1 $ [see figure~\ref{t-fig4}(a)], but, for $\Omega > 1$, the modes become stable. Upon examining the other two branches, namely, the first-excited branch and the second-excited branch, we observe that instability gain becomes finite only for $\Omega < 0$ as illustrated in Fig.~\ref{t-fig4}(b), and Fig.~\ref{t-fig4}(c), respectively. Interestingly, the low-lying and first-excited branches of the eigenspectrum display multiple instability bands for $\Omega < 0$, whereas the second-excited branch exhibits only a single instability band starting at $\Omega \lesssim -1.66$. Upon considering relatively stronger SOC strength ($k_{L} = 4.0$),  multi-band instabilities appear in both the low-lying and the first-excited branches [see Figs.~\ref{t-fig4}(d)-(f)]. The second-excited branch, however, still only shows a single instability band, similar to the case with $k_L = 1$. Comparing the instability features across the range of Rabi frequencies from $-\Omega$ to $+\Omega$, we observe a transition from multi-band instability to a single instability band in the low-lying branch, accompanied by the existence of a primary band in the first-excited branch for $\Omega > 0$. The primary instability band in the low-lying branch appears at $k_L^2 = \Omega$. In the second-excited branch, a single instability band is present for $\Omega \lesssim -1.44$. Furthermore, the instability gains ($\Sigma_{ll}$, $\Sigma_{fe}$, $\Sigma_{se}$) in all three branches increase as the Rabi coupling strength decreases, and they exhibit symmetry about $q_x$. 


Overall, we find that SOC BECs with FM interaction exhibit a single UAC between $\omega_{ll}$ and $\omega_{fe}$, as previously reported~\cite{Gangwar2024}. This UAC emerges at critical values of coupling strength, following the relation $\Omega \approx 0.1278 k_{L}^{2} - 1.1358$, with its origin at $(k_{L}, \Omega) = (2.94, 0.01)$. Interestingly, in MI analysis~\cite{Robins2001, Limod2017}, we observe that for $\Omega < 0$, the first excited branch ($\omega_{fe}$) exhibits a second UAC with the second excited branch ($\omega_{se}$), accompanied by the emergence of more number of multi-band instabilities compared to cases where $\Omega > 0$.



After analyzing the collective excitation for FM interaction, we now present the corresponding analysis for the AFM interaction in the next section.

\subsection{Collective excitation spectrum of AFM SOC-BECs ($c_{0} > 0$, $c_{2} > 0$)}
\label{sec:5}

\begin{figure}[!htp]
\begin{centering}
\centering\includegraphics[width=0.99\linewidth]{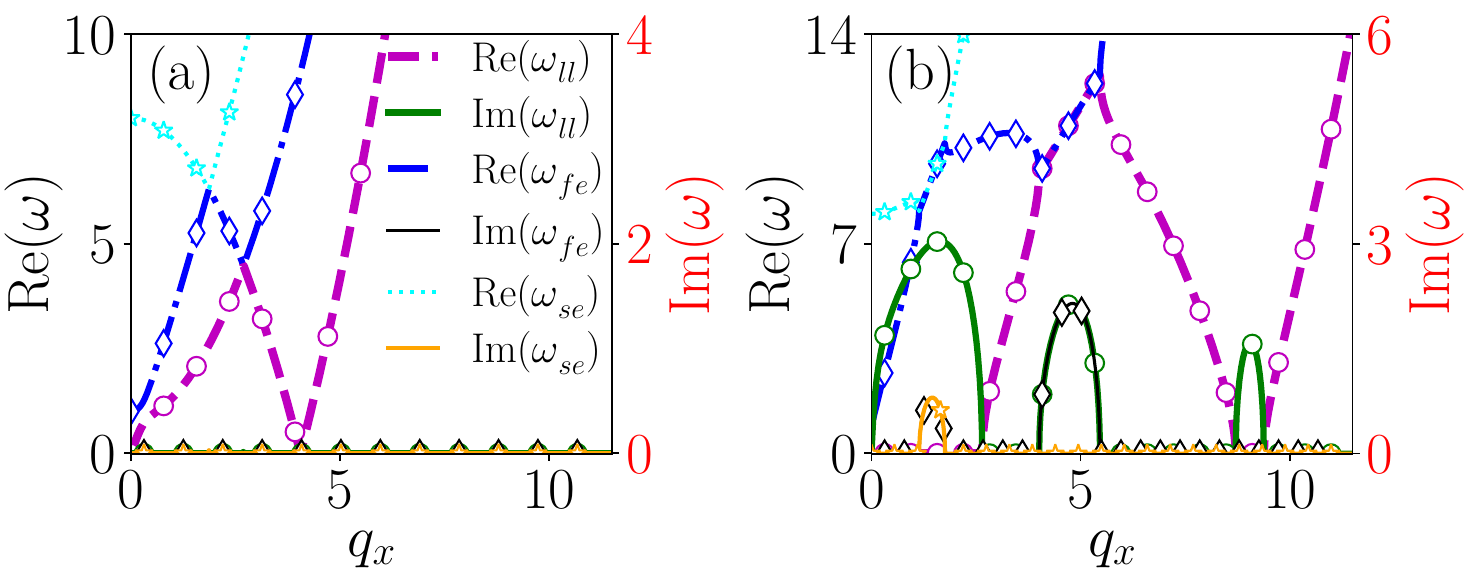}
\caption{Eigenspectrum with AFM interaction ($c_{0} = 5.0$, $c_{2} = 5.0$) for (a) ($k_{L}, \Omega$) = ($0.5, 1.0$), and (b) ($k_{L}, \Omega$) = ($4.5, 1.0$). The lines and symbols representation is same as in Fig.~\ref{t-fig0}(a i) and (b i). 
In panel (a), a stable avoided crossing appears between first-excited ($\omega_{fe}$) and second-excited ($\omega_{se}$) as well as low-lying ($\omega_{ll}$) and $\omega_{fe}$ branches. As SOC is increased in panel (b) multi-band instability emerges in the eigenspectrum accompanied by the appearance of UAC  between $\omega_{fe}$- $\omega_{se}$ and $\omega_{ll}$- $\omega_{fe}$ branches. The first-excited branch exhibits the double UAC. The right vertical axis represents the imaginary eigenfrequencies.}  
\label{t-fig00}
\end{centering}
\end{figure}

In this section, we present the collective excitation spectrum for AFM interaction SOC-BECs by considering $c_0 = c_2 = 5.0$. In regime I, with $k_L = 0.5$ and $\Omega = 1.0$, the eigenspectrum exhibits only real eigenfrequencies. However, avoided crossings occur between the $\omega_{ll}-\omega_{fe}$ branches at $q_x \sim 2.70$ and the $\omega_{fe}-\omega_{se}$ branches at $q_x \sim 1.87$ [see Fig.~\ref{t-fig00}(a)]. As the imaginary part of the eigenspectrum is zero ($\mathrm{Im}(\omega_j) = 0$), the condensate remains dynamically stable. The involvement of the first-excited branch in both of these stable avoided crossings results in a double stable avoided crossing near $q_x \sim 1.87$ and $q_x \sim 2.70$, as depicted in Fig.~\ref{t-fig00}(a). Furthermore, the minimum of the low-lying excitation spectrum occurs at $q_x = 4.06$, corresponding to the phonon energy minimum. This minimum corresponds to the rotonic minimum and is responsible for the emergence of the plane-wave phase in spin-1 AFM SOC BECs~\cite{Martone2012, Yu2016}. Such features are not observed in FM SOC BECs.

In Fig.~\ref{t-fig00}(b), we show the eigenspectrum corresponding to the regime II where we have considered $k_{L} = 4.5$ and $\Omega = 1.0$. The other parameters are the same as those in panel (a). In this regime, we notice the appearance of multi-band imaginary eigenfrequency both in the $\omega_{ll}$ and $\omega_{fe}$ branches. There are three instability bands present in the low-lying branch whose location and corresponding amplitude are given as $\{q_{x}, \omega\}=\{1.63, 3.037\}, 
\{4.81, 2.149\}, \{9.06, 1.585\}$. However, the other two instability bands appear in the first excited branch with position and amplitudes $\{q_{x}, \omega\}=\{1.46, 0.802\}, \{4.81, 2.149\}$. Additionally, we notice the presence of a single-band instability in the $\omega_{se}$ branch with $\{q_{x}, \omega\}=\{1.46, 0.802\}$ which arises solely due to crossing between $\omega_{se}-\omega_{fe}$. The presence of imaginary eigenfrequency leads to dynamical instability in regime II, as also obtained in~\cite{Goldstein1997, Ozawa2013, Mishra2021}. The $\omega_{fe}$  branch experiences a double UAC, arising from avoided crossings with the other two branches. The first UAC between $\omega_{ll}-\omega_{fe}$ appears in the quasimomentum range $4.01 \lesssim q_{x} \lesssim 5.49$, while, the second UAC appears between $\omega_{fe}-\omega_{se}$ in the quasi-momentum range $1.14 \lesssim q_{x} \lesssim 1.78$ [see figure~\ref{t-fig00}(b)]. Such an instability is responsible for a $I_{o}$ type of dynamical instability~\cite{Bernier2014, Pu2016}, which will be discussed in the later part of the paper. 

\begin{figure*}[!htb]
\begin{centering}
\centering\includegraphics[width=\linewidth]{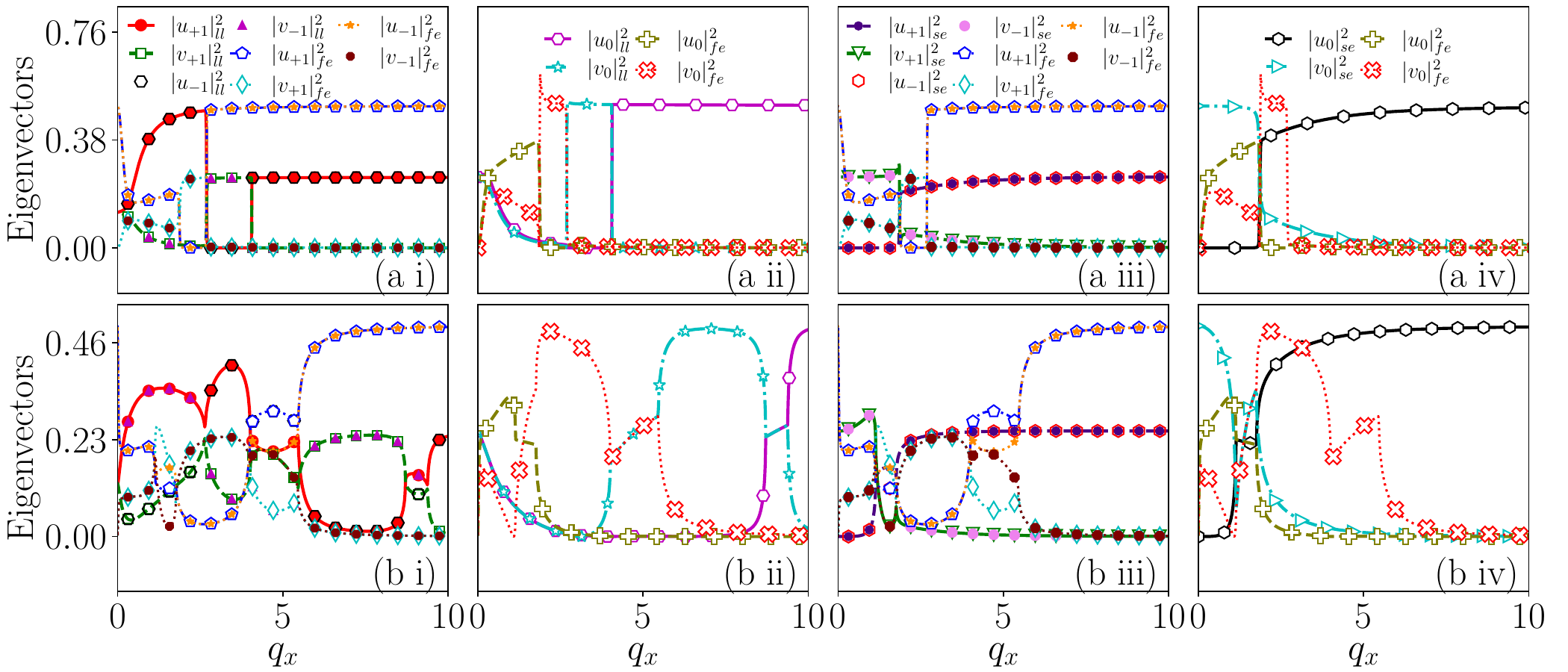}
\caption{Eigenvectors corresponding to Figs.~\ref{t-fig00}(a) and  (b), in (a) and (b), respectively. (a i, a ii), and (b i, b ii) represents eigenvectors of the low-lying and first-excited branch, while (a iii, a iv) and (b iii, biv) for the second-excited and first-excited branch of the eigenspectrum. The eigenvectors of the low-lying branch are given as $\vert u_{+1}\vert^{2}_{ll}$ (red dot), $\vert u_{-1}\vert^{2}_{ll}$ (black hexagons), $\vert u_{0}\vert^{2}_{ll}$ (magenta hexagons), $\vert v_{0}\vert^{2}_{ll}$ (cyan open stars), $\vert v_{+1}\vert^{2}_{ll}$ (green open squares), and $\vert v_{-1}\vert^{2}_{ll}$ (magenta triangles). For the first-excited branch, $\vert u_{+1}\vert^{2}_{fe}$ (blue open pentagon), $\vert u_{-1}\vert^{2}_{fe}$ (orange stars), $\vert u_{0}\vert^{2}_{fe}$ (olive open plus), $\vert v_{0}\vert^{2}_{fe}$ (red open x with the dotted line), $\vert v_{+1}\vert^{2}_{fe}$ (cyan open diamond), and $\vert v_{-1}\vert^{2}_{fe}$ (maroon dots). The eigenvectors of the second-excited branch are given as $\vert u_{+1}\vert^{2}_{se}$ (indigo dot), $\vert u_{-1}\vert^{2}_{se}$ (red hexagons), $\vert u_{0}\vert^{2}_{se}$ (black hexagons), $\vert v_{0}\vert^{2}_{se}$ (cyan triangle left), $\vert v_{+1}\vert^{2}_{se}$ (green open triangle down), and $\vert v_{-1}\vert^{2}_{se}$ (violet dots). (a i - a iv) depict in-phase behaviour (density-like mode) with the occurrence of flip at the point of stable avoided crossing between $\omega_{ll} - \omega_{fe}$, and $\omega_{se} - \omega_{fe}$. (b i, b iii) exhibit the out-of-phase behaviour (spin-like mode). At the point of UAC, eigenvectors are out-of-phase among the branches. However, (b ii, b iv) only depict density-like mode.} 
\label{t-fig01}
\end{centering}
\end{figure*}
Next, we analyze the corresponding eigenvectors of the regime I and II for the AFM SOC BECs. In Fig.~\ref{t-fig01}, we demonstrate the nature of eigenvectors corresponding to Fig.~\ref{t-fig00}. First, we report for regime I the respective eigenspectrum is presented in Fig.~\ref{t-fig00}(a). As all the eigenfrequencies  $\omega_{ll}$, $\omega_{fe}$ and $\omega_{se}$ are real, the corresponding eigenvector components display in-phase (density-like) behaviour in quasi-momentum direction ($q_x$), which holds the criterion given in Eq.~\ref{eq:density} (also holds for the eigenvectors of the second-excited branch). A flip in the eigenvector components occurs at the point of first stable avoided crossing around $q_x\sim 2.70$ between the $\omega_{ll}-\omega_{fe}$ branches [see Figs.~\ref{t-fig01}(a i)] similar as reported in the ref~\cite{Abad2013}. As the second stable avoided crossing is observed between $\omega_{fe}-\omega_{se}$, the eigenvector components for corresponding branches also show the flipping tendency at the point of crossing at $q_x\sim 1.87$ [see Figs.~\ref{t-fig01}(a iii)]. The zeroth component of eigenvectors of low-lying, first- and second-excited branches of the eigenspectrum show density-like mode, which is given in Figs.~\ref{t-fig01}(a ii), and (a iv). Overall, we observe that at the point of stable avoided crossing between the branches, the flip in eigenvectors of both branches occurs simultaneously at $q_{x} \approx 2.70$ and $ \approx 1.87$. Moreover, further flip in the eigenvectors of the low-lying branch takes place at $q_{x} \approx 4.06$, when the eigenvalue spectrum approaches zero, where $\text{Re}(\omega) = \text{Im}(\omega) = 0$. 

We present the nature of eigenvectors for high $k_L$ corresponding to the eigenspectrum shown in Fig.~\ref{t-fig00}(b). We observe that due to the presence of complex eigenfrequency ($\mathrm{Im}(\omega_j) \neq 0$), the eigenvector components show the spin-like (out-of-phase) behaviour characterized by the Eq.~\ref{eq:spin} (the same applied for the eigenvectors of the second-excited branch). As the low-lying and first-excited branches have multi-band instability in the eigenspectrum, the eigenvectors exhibit a transition from spin-like to density-like mode resulting in a mixed mode. At the first UAC between $\omega_{ll}-\omega_{fe}$ branches occur in the quasi-momentum range $4.01 \lesssim q_{x} \lesssim 5.49$ in the eigenspectrum [see Fig.~\ref{t-fig00}(b)], the corresponding eigenvectors exhibit complicated out-of-phase behaviour among these branches as illustrated in Fig.~\ref{t-fig01}(b i). As we look at the eigenvector in the second UAC region ($1.14 \lesssim q_{x} \lesssim 1.78$), we find that they exhibit similar out-of-phase features [see Fig.~\ref{t-fig01}(b iii)]. Interestingly, the zeroth component of eigenvectors for the low-lying, first- and second-excited branches exhibit density-like mode independently [see Fig.~\ref{t-fig01}(b ii), and (b iv)]~\cite{Gangwar2024}.
In the preceding section, we derived the multi-band imaginary eigenfrequencies for the low-lying and first-excited branches, as well as the single-band imaginary eigenfrequency for the second-excited branch of the spectrum. The instability gains in these eigenspectra are characterized as $\Sigma_{ll, fe, se} = \left\vert \mathrm{Im}(\omega_{ll, fe, se}) \right\vert$.
\begin{figure}
\begin{centering}
\centering\includegraphics[width=0.99\linewidth]{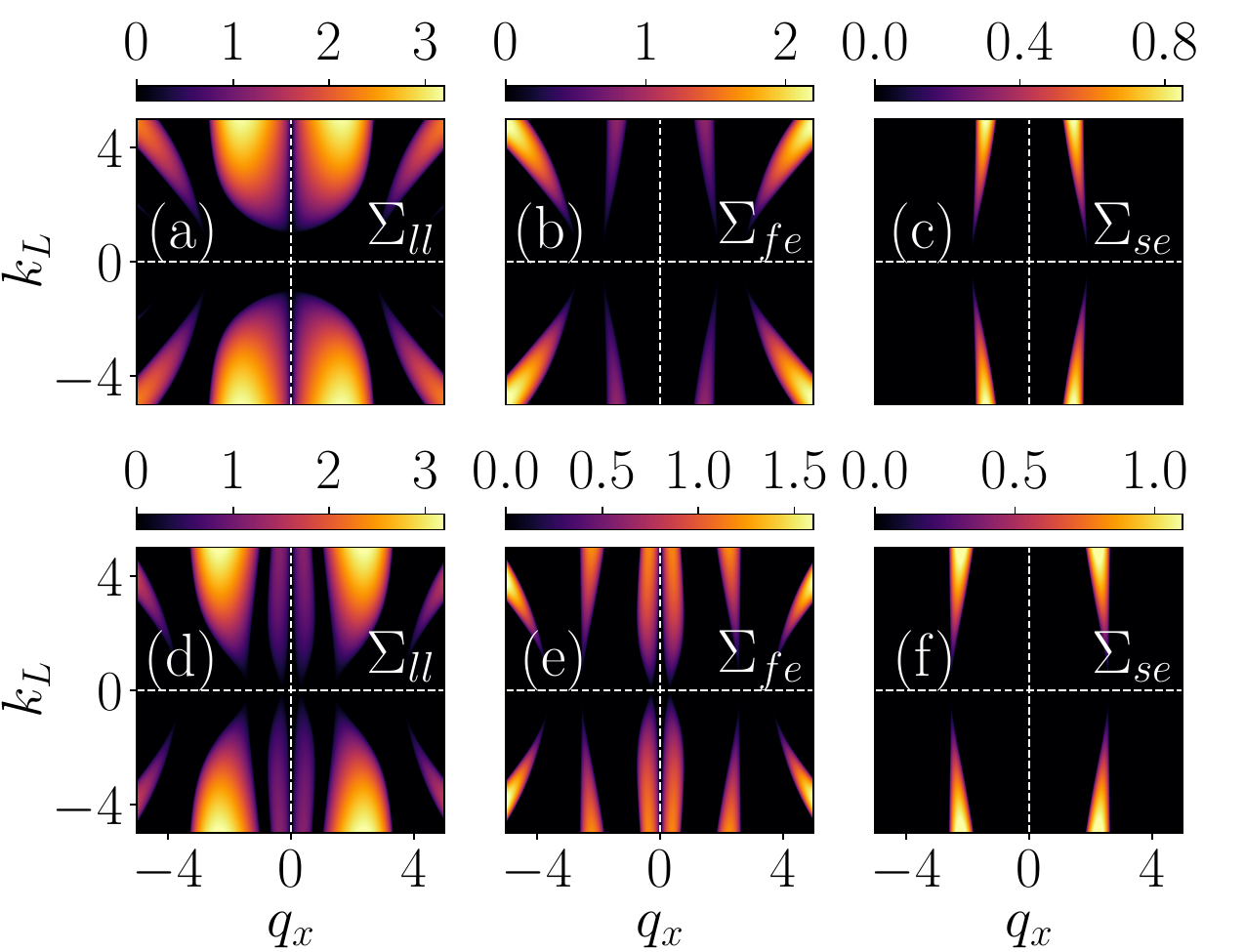}
\caption{{MI gain in the $k_{L} - q_{x}$ phase plane for (a-c) $\Omega = 1$ and (d-f)  $\Omega = -1$. The interaction strengths are $c_{0} = c_{2} = 5.0$. 
In the panel, the pseudo color represents similar denotations as Fig.~\ref{t-fig4}.  In the top row, we obtain the instability bands at $k_{L} = 1.0$, following the critical relation $k_{L}^{2} = \Omega$,  while in the bottom row, it appears for $k_{L} > 0.25$. Symmetry arguments are similar to Fig.\ref{t-fig2}.}}
\label{t-fig1}
\end{centering}
\end{figure}%
After analyzing the presence of instability in the collective excitation spectrum of AFM SOC-BECs, we aim to investigate the nature of this instability through instability gains using phase plots, varying the parameters $k_L$ and $\Omega$. We begin by examining the role of $k_L$ (SOC strength) on the instability gain of AFM condensates with interaction strengths $c_0 = c_2 = 5.0$. In Figs.~\ref{t-fig1}(a)-(c), the Rabi coupling strength is fixed at $\Omega = 1$. Instability emerges when $k_L > 1$. In the low-lying branch ($\Sigma_{ll}$), we observe the emergence of double-band instability along the quasi-momentum direction. The growth of the primary instability increases with increasing SOC strength. Additionally, the bandwidth of the secondary instability widens, exhibiting a horn-like shape. The first- and second-excited branches also display instability gains ($\Sigma_{fe}$ and $\Sigma_{se}$), characterized by double- and single-band structures, respectively. For the instability gain of the first excited state ($\Sigma_{fe}$), two humps appear: the primary instability band remains constant, while the secondary instability band expands with increasing SOC strength. The instability gain of the second-excited branch ($\Sigma_{se}$) and its bandwidth also exhibit an increasing trend as $k_L$ increases.
Interestingly, for a negative Rabi coupling ($\Omega = -1$), we observe a distinctly different trend compared to the positive Rabi coupling case. In a spin-1 system, negative Rabi coupling injects additional energy into the system~\cite{Ravisankar2020}, resulting in a higher number of instability bands. Consequently, instability arises earlier (at $k_L > 0.25$) compared to the positive Rabi coupling case [see Figs.~\ref{t-fig1}(d)-(f)]. Here, triple instability bands appear in the low-lying and first-excited branches. The first band remains constant, while the bandwidth and instability gain of the second band varies with $k_L$, and the third band grows gradually with increasing SOC strength, consistent with the previous case. For both $\Omega = 1$ and $\Omega = -1$, the instability bandwidth of all branches remains symmetric about the quasi-momentum $q_x$ and the SOC strength $k_L$.
\begin{figure}
\begin{centering}
\centering\includegraphics[width=0.99\linewidth]{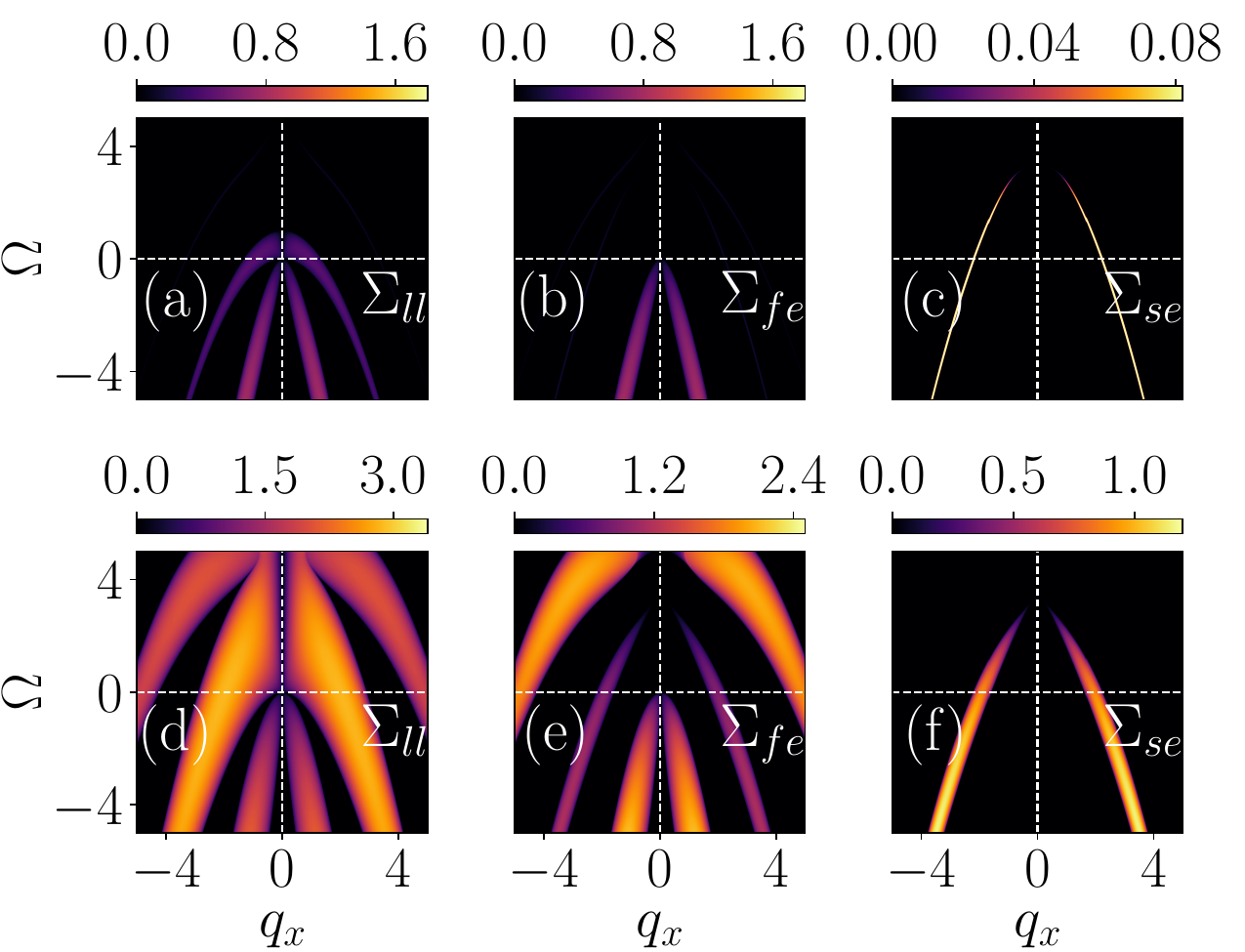}
\caption{{MI gain in $\Omega - q_{x}$ phase diagram for (a-c) $k_{L} = 1$, and (d-f) $k_{L} = 4.0$. Here, the density-density interaction, $c_{0} = 5.0$, and spin-dependent interaction $c_{2} = 5.0$. The pseudo-color bar in the figure is similar to Fig.~\ref{t-fig4}.
Multi-band instability appears for the low-lying and first-excited branches of the spectrum, while the second-excited branch shows single-band instability. The instability gain appears is not symmetric about $\Omega$ and increases upon decreasing the $\Omega$. However, it shows symmetric behaviour about $q_{x}$.}}
 \label{t-fig3}
\end{centering}
\end{figure}%
Next, we investigate the effect of Rabi coupling strength on the instability gain, denoted as $\Sigma_{ll, fe, se}$, in spin-orbit (SO) coupled spin-1 BECs with AFM interactions, where $c_0 = c_2 = 5$. In Figs.~\ref{t-fig3}(a)–(c), we fix the SO coupling strength at $k_L = 1$ and vary the Rabi coupling strength, $\Omega$. This reveals double instability bands in the low-lying branch of the eigenspectrum, whereas the first-excited and second-excited branches exhibit only single-band instability. In the low-lying branch, the primary instability band emerges at $\Omega = 1.0$, increases in magnitude for $\Omega < 1.0$, and disappears, becoming stable for $\Omega > 1.0$. Additionally, a secondary instability band appears for $\Omega < 0$. In the first-excited branch, instability gain is observed for $\Omega < 0$, inheriting an unstable avoided crossing (UAC) between the low-lying and first-excited branches. In the second-excited branch, a single instability band emerges, growing for $\Omega \lesssim 3.15$; at $\Omega < 0$, a second UAC appears. Thus, when $\Omega < 0$, the AFM system exhibits double UACs and double instability gain bands [see Fig.~\ref{t-fig3}(c)]. However, the amplitude of the instability gain in this branch is significantly smaller compared to the other two branches.

For a relatively strong SO coupling strength of $k_L = 4.0$, we observe multi-band instability in the low-lying and first-excited branches of the eigenspectrum, while the second-excited branch displays single-band instability for $\Omega < 3.0$. Due to the significantly higher SO coupling strength, the instability gain exhibits both a larger amplitude and higher coverage in the phase plane [see Figs.~\ref{t-fig3}(d)–(f)]. Comparing the two cases, we find that for weak SO coupling ($k_L = 1$), the maximum unstable phase occurs only for negative $\Omega$ (i.e., $-\Omega$), with stability observed for positive $\Omega$ (i.e., $+\Omega$). In contrast, for strong SO coupling ($k_L = 4.0$), the entire considered range of $\Omega$ is unstable.

In Figs.~\ref{t-fig3}, the instability gain along the quasi-momentum direction ($q_x$) increases as the Rabi coupling strength decreases, displaying symmetric behaviour about $q_x$. Furthermore, we confirm that in regime II of the AFM interaction, double UACs are present for both positive and negative $\Omega$ ($\pm \Omega$), whereas, in the ferromagnetic (FM) interaction, double UACs occur only for negative $\Omega$ ($-\Omega$). Thus, both systems exhibit double UACs, with their appearance dependent on the Rabi coupling strength.
\begin{figure}
\begin{centering}
\centering\includegraphics[width=0.9\linewidth]{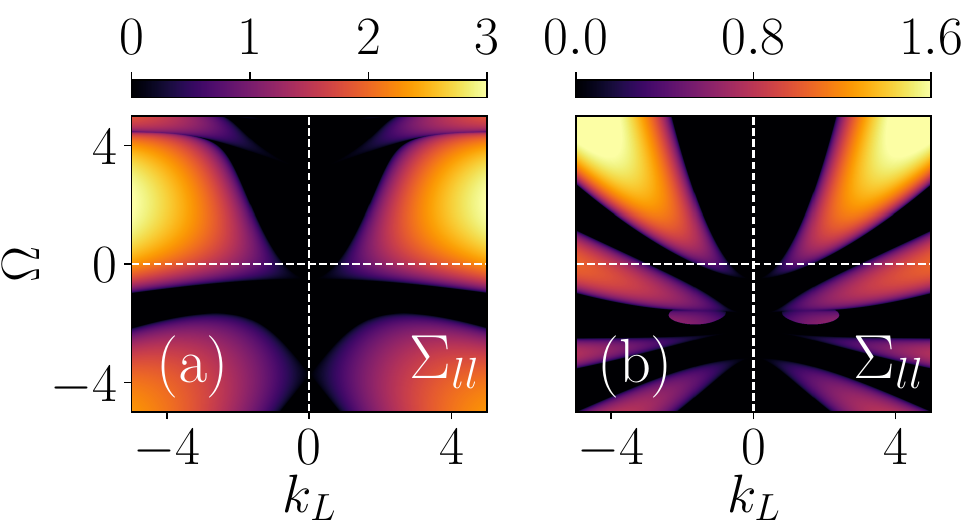}
\caption{{Pseudo color representation of the MI gain for $\mathrm{\Sigma}_{ll} \equiv \mathrm{\vert Im(\omega}_{ll}) \vert$ in $k_L-\Omega$ phase plane for (a) ($c_{0},c_2$) = ($5, 5$) and (b) ($c_{0},c_2$) =($5,-2$) at quasi-momentum $q_x = 1$.  Multi-band instability appears in each case, which increases upon increasing the $k_{L}$. The instability bands show symmetric behaviour about $k_{L}$, however, it is not symmetric with the variation in $\Omega$.}}
 \label{t-fig5}
\end{centering}
\end{figure}%

Further, we study the behaviour of the system at fixed interaction strengths $c_{0}$, $c_{2}$, and quasi-momentum $q_{x}$, upon simultaneous varying the SO coupling ($k_{L}$) and Rabi coupling ($\Omega$) strengths [see Figs.~\ref{t-fig5}(a), (b)]. We demonstrate instability gain of the low-lying branch of the eigenspectrum corresponding to two different sets of interaction strengths at $q_{x} = 1$. In Figs.~\ref{t-fig5}(a), we consider interaction strengths corresponding to AFM interaction, and in figure~\ref{t-fig5}(b), corresponding to FM interaction. We obtain multi-band instability for both cases, which is not symmetric about $\Omega$. The instability gain is symmetric about $k_{L}$ and increases upon increasing the SO coupling strength. Here, we conclude that the regime I, ($\Omega > k_{L}^{2}$) is dynamically stable while the regime II ($\Omega < k_{L}^{2}$) is dynamically unstable and exhibit multi-instability bands. Moreover, with respect to the negative Rabi couplings, both spinor BECs exhibit double UAC. Additionally, we observe that the instability gain of the FM condensate is reduced by half for AFM condensate, indicating that the latter is dynamically more unstable and sensitive to perturbations.  
\section{Impact of density-density interaction and spin-exchange interaction on the MI}
\label{sec:6}
So far we have analyzed the effect of $k_{L}$ and $\Omega$ on the collective excitation spectrum for the FM ($c_{0}, c_{2}$) = ($5.0, -2.0$)  and AFM ($c_{0}, c_{2}$) = ($5.0, 5.0$) interaction phases of the SOC spin-1 condensate. For FM interactions, the instabilities mainly appear in the low-lying and first-excited branches of the eigenspectrum. However, for AFM interactions, they appear in the low-lying, first- and second-excited branches. Next, to make our analysis more general, we scan a wide range of interactions ($(c_{2}, c_{0})\in [-10,10]$) for low SOC ($k_{L}^{2} < \Omega$) and high SOC ($k_{L}^{2} > \Omega$) regimes. 

\begin{figure}
\begin{centering}
\centering\includegraphics[width=0.99\linewidth]{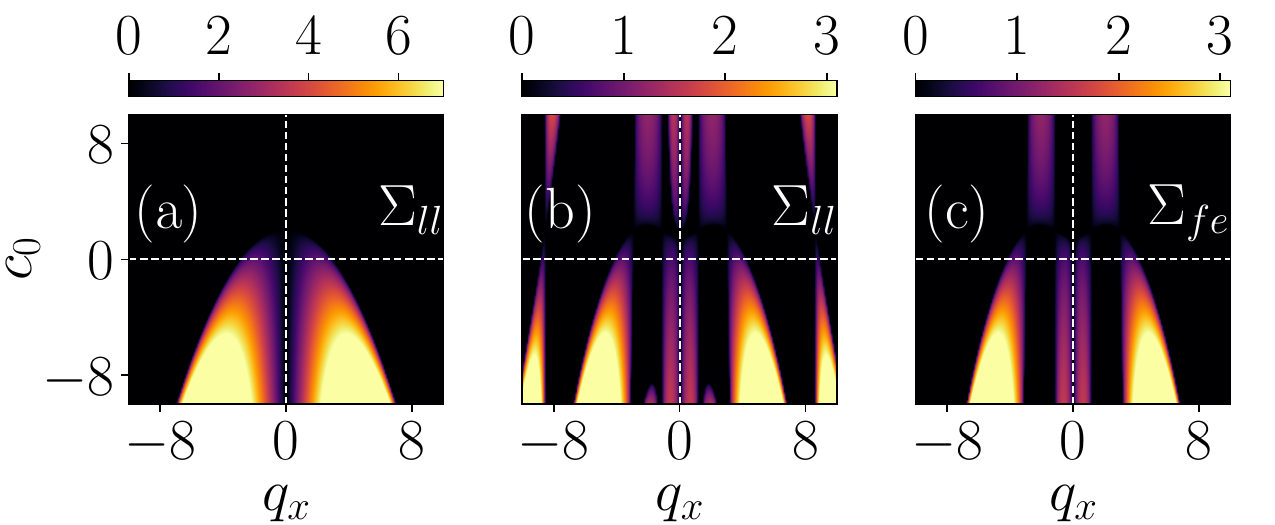}
\caption{{(a) Pseudo color representation of MI gain for low-lying mode $\mathrm{\Sigma}_{ll} \equiv \mathrm{\vert Im(\omega}_{ll}) \vert$  in $c_{0} - q_{x}$ phase plane with 
 ($k_{L}, \Omega$) = ($0.5, 1.0$).  (b) and (c) represent the pseudo color representation of MI gain for low-lying mode $\mathrm{\Sigma}_{ll} \equiv \mathrm{\vert Im(\omega}_{ll}) \vert$  and  first excited mode $\mathrm{\Sigma}_{fe} \equiv \mathrm{\vert Im(\omega}_{fe}) \vert$ respectively in the $c_{0} - q_{x}$ phase plane for  ($k_{L}, \Omega$) = ($4.5, 1.0$). The spin-exchange interaction is $c_{2} = -2.0$.  In (a), the instability gain appears only in the low-lying branch, while in (b, c), it appears in a low-lying and first-excited branch of the excitation spectrum. In the panel, the instability gain increases upon decreasing the $c_{0}$, showing symmetric behaviour about $q_{x}$. However, it is not symmetric about $c_{0}$.}}
\label{t-fig6}
\end{centering}
\end{figure}

\begin{figure}
\begin{centering}
\centering\includegraphics[width=0.99\linewidth]{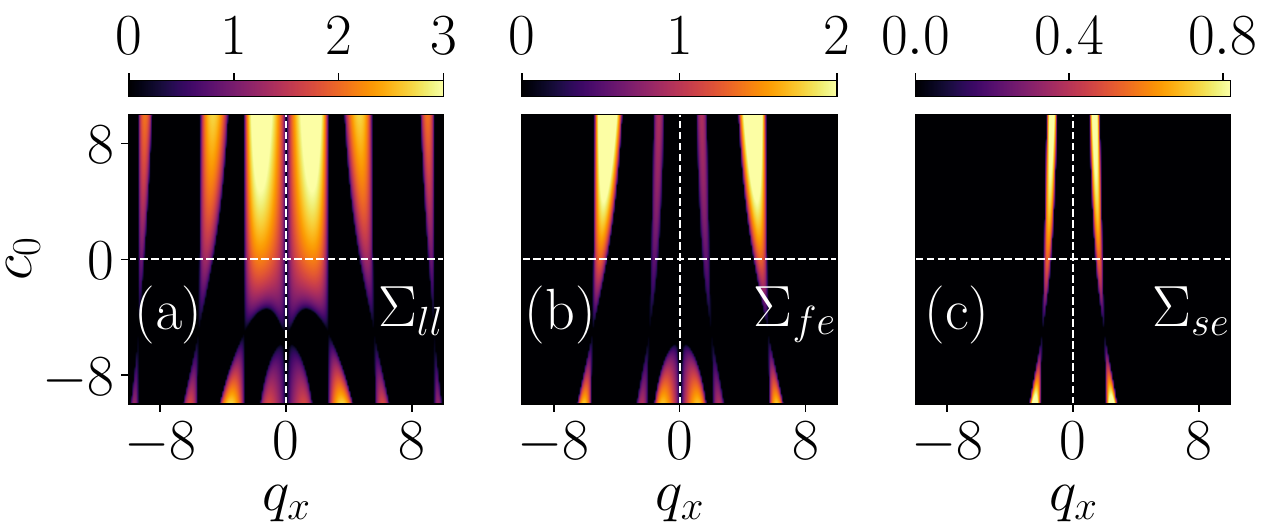}
\caption{{MI of AFM interaction ($c_{2} = 5.0$) in $c_{0} - q_{x}$ phase for (a-c) $k_{L} = 4.5, \Omega = 1.0$. The pseudo color bar is similar to Fig.~\ref{t-fig4}.
Multi-band instability in low-lying and first-excited branches and single-band instability in the second-excited branch of eigenspectrum increases upon increasing the $c_{0}$ and symmetry arguments are similar to figure~\ref{t-fig6}.}}
 \label{t-fig6a}
\end{centering}
\end{figure}

%
First, we consider the effect of density-density interaction term $c_{0}$ by choosing $c_{2}$ positive or negative. At first, we choose $c_{2} =-2.0$ along with two different sets of coupling strengths ($k_{L}$, $\Omega$). For ($k_{L}^{2} < \Omega$), we find that the instability gain is zero as long as the total interaction strength remains repulsive, i.e., $c_{0} + c_{2} > 0$. However, single-band instability appears in the low-lying branch of the eigenspectrum when the total interaction strength is attractive, i.e.,  $c_{0} + c_{2} < 0$. For $c_{2} = -2.0$ and $c_{0} > 2.0$, the instability gain is absent. The instability appears where the interaction follows the relation $c_{0} + c_{2} < 0$. The instability gain appears to increase upon further decreasing the $c_{0}$ [see figure~\ref{t-fig6}(a)]. However, there is a lack of instability in the first- and second-excited branches (not shown here). Now, we consider regime II that holds $k_{L}^{2} > \Omega$, and we obtain multi-band instability that appears in the low-lying and first-excited branches of the spectrum. The instability gain of the spectrum increases upon decreasing the $c_{0}$ [see Figs.~\ref{t-fig6}(b), and (c)]. On the other hand, instability gain is absent in the second-excited branch of the eigenspectrum. We found that regime I has instability only for the attractive system and no UAC, but regime II is stable only for $ c_{0} + c_{2} \approx 0$, otherwise exhibits UAC, also no appearance of double UAC. For both sets of coupling strengths, the instability gain of the spectrum is symmetric about $q_{x}$ and not symmetric about $c_{0}$. 
In the above part, we analyzed the effect of $c_{0}$ in the presence of $c_{2} = -2.0$. Further, to analyze the impact of positive interaction, we consider $c_{2} = 5.0$. Here also we consider the low-$k_L$ regime, regime I ($k_{L}^{2} < \Omega$) and high $k_L$ regime,  regime II ($k_{L}^{2} > \Omega$). For regime I, we consider $c_{0}$ in the range [-5:5], and obtain only real eigenfrequencies in the eigenspectrum. Therefore, the instability gain remains absent and is responsible for the dynamical stable phases (not shown here). Further, we choose regime II, which exhibits multi-band instability that appears in low-lying and first-excited branches of the eigenspectrum [see Figs.~\ref{t-fig6a}(a) and (b)], while the second-excited branch shows single band instability [see figure~\ref{t-fig6a}(c)]. 
The regime II is stable only for the attractive $c_0$ interactions results as $c_{0} + c_{2} \approx 0$. The instability gain increases upon increasing $c_{0}$, showing symmetric behaviour about $q_{x}$, while not symmetric about $c_{0}$.%

\begin{figure}
\begin{centering}
\centering\includegraphics[width=0.99\linewidth]{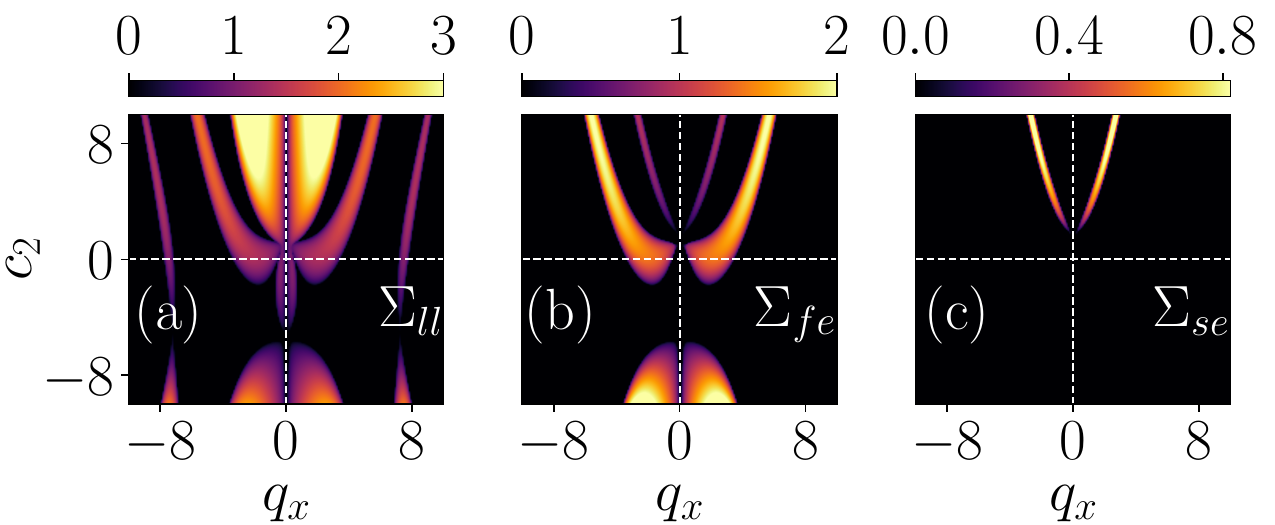}
\caption{{MI in $c_2- q_{x}$ phase plane for (a-c) $k_{L} = 4.0, \Omega = 1.0$, with $c_{0} = 5.0$. The pseudo color bar is similar to Fig.~\ref{t-fig4}.
The instability gain of the spectrum increases upon increasing $c_{2}$, showing the symmetric behaviour about $q_{x}$. However, it is not symmetric about $c_{2}$.}}
 \label{t-fig7}
\end{centering}
\end{figure}

After exploring the density-density interaction parameter regime, we now shift our focus to analyzing the impact of the spin-exchange interaction strength, $c_2$, on the instability while keeping the density-density interaction strength fixed at $c_0 = 5.0$. In regime I, we consider $c_{2}$ in the range [-5:5], and the eigenspectrum yields only real eigenfrequencies, indicating an absence of instability (not shown here). In contrast, for regime II, we observe the emergence of multi-band instability in both the low-lying and first-excited branches of the eigenspectrum. The low-lying branch remains unstable across the considered range of $c_2$ (except where $c_0 + c_2 \approx 0$), while the first-excited branch exhibits instability accompanied by the first UAC. Conversely, the second-excited branch displays single-band instability when the effective interaction is AFM, coinciding with the appearance of a second UAC in the system, but only for $c_2 > 2$ [see Figs.~\ref{t-fig7}(a)-(c)]. The instability gain increases upon increasing the $c_{2}$, showing symmetric behaviour about $q_{x}$, preserving asymmetric behaviour about $c_{2}$.

\begin{figure}[h]
\begin{centering}
\centering\includegraphics[width=0.99\linewidth]{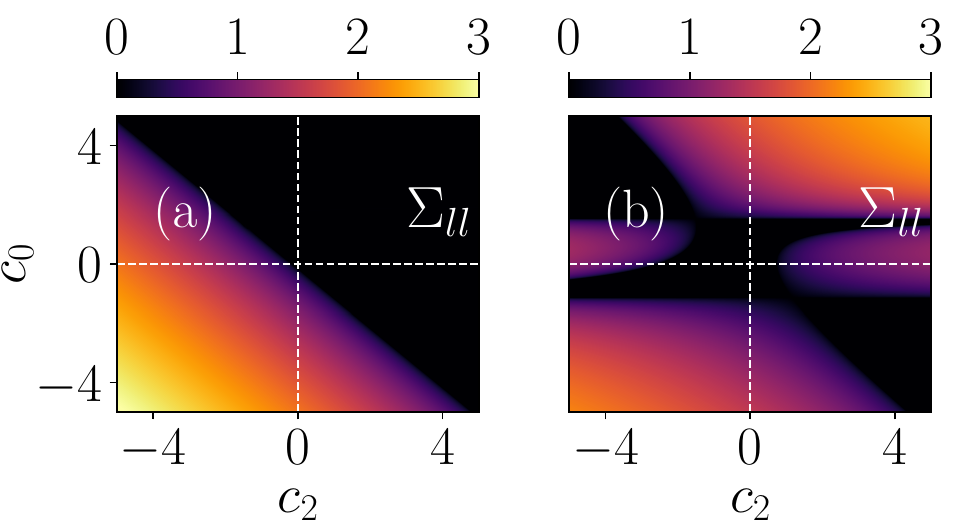}
\caption{{The pseudo color representation of MI gain of lowest eigen mode $\mathrm{\Sigma}_{ll} \equiv \mathrm{\vert Im(\omega}_{ll}) \vert$ in $c_0-c_2$ phase plane for (a) ($k_L, \Omega$) =($0.5, 1.0$) and (b) ($k_L, \Omega$) =($4.0, 1.0$). The quasi-momentum strength is fixed at $q_{x} = 1$. In (a) the instability gain appears only in the low-lying branch when the total interaction strength is attractive $c_{0} + c_{2} < 0$, while in (b) the spectrum  exhibits multi-band instability. The instability gain in the eigenspectrum is neither symmetric about $c_{0}$ nor about $c_{2}$.}}
 \label{t-fig8}
\end{centering}
\end{figure}%


In Fig.\ref{t-fig8}(a), we show the instability gain of the eigenspectrum upon simultaneously varying density-density interaction strength ($c_{0}$) and spin-exchange interaction strength ($c_{2}$) for  ($k_L, \Omega$) =($0.5, 1.0$). We find that the instability only occurs for the low-lying branch of the eigenspectrum for attractive interaction when $c_{0} + c_{2} < 0$. However for the repulsive case ( $c_{0} + c_{2} > 0$) system appears to be stable one. In Fig.\ref{t-fig8}(b), we show the instability gain corresponding to the low-lying eigenspectrum for $k_{L} =4.0$ and  $\Omega=1.0 $). For this case, we find the presence of a wider instability region in the entire range of interaction strengths, including both attractive and repulsive, as well as for the mixed case. Interestingly, we find that, for the mixed case of interactions, the system gets stabilized. Overall, we find that for the repulsive case, regime II shows instability, however, for regime I, the repulsive interaction shows a stable nature. The instability gain of the eigenspectrum is neither symmetric about $c_{0}$ nor symmetric about $c_{2}$.%

\begin{figure}
\centering\includegraphics[width=0.99\linewidth]{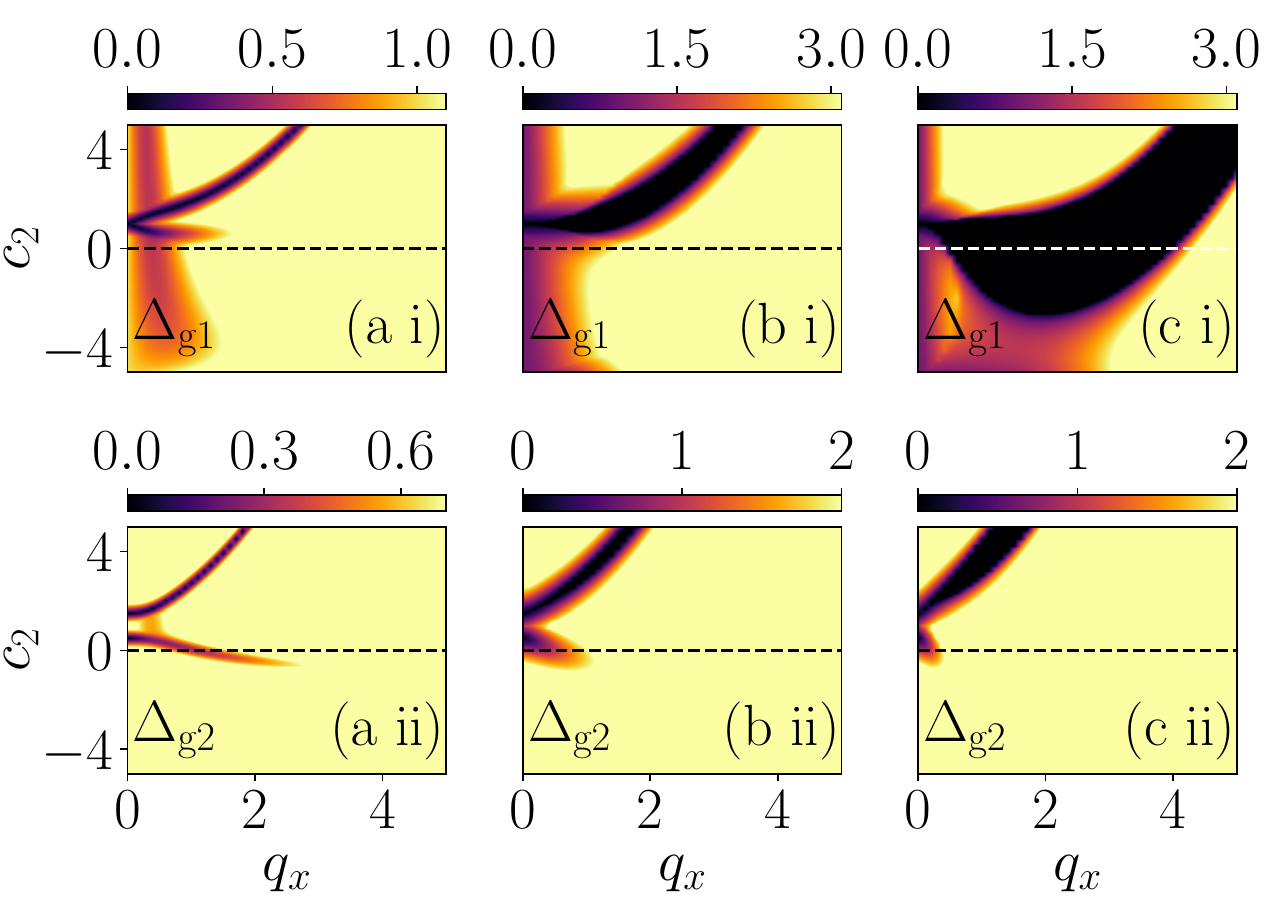}
\caption{Pseudo color representation of the change in band gap $\mathrm{\Delta_{g1}} = \mathrm{\omega}_{fe} - \mathrm{\omega}_{ll}$ and  $\mathrm{\Delta_{g2}} = \mathrm{\omega}_{se} - \mathrm{\omega}_{fe}$  for different $k_L$ in $c_{2}-q_{x}$ plane by keeping $\Omega = 1.0$. (a i, a ii) for $k_L=0.5$, (b i, b ii) for $k_L= 2.0$ and (c i, c ii) for $k_L=4.5$.
For (a i, b i), the band gap $\Delta_{g1}$ closes for positive $c_{2} > 0$ (AFM interaction). However, in (c i), it closes for both AFM ($c_{2}>0$) and FM ($c_{2}<0$) interactions. In (a ii - c ii), the band gap $\mathrm{\Delta_{g2}}$ closes for AFM interactions only ($c_{2}>0$).
} 
\label{fig82} 
\end{figure} 
\section{Characterization of double unavoided crossings using the band gap $\mathrm{\Delta_{g1}}$, and $\mathrm{\Delta_{g2}}$ of the eigen-spectrum}
\label{sec:7}
After gaining a comprehensive understanding of the instability of FM and AFM interactions, we now characterize the band gap between the eigenbranches. The band gaps in the eigenspectrum are defined as the difference between the first-excited and low-lying states, expressed as $\mathrm{\Delta_{g1}} = \omega_{fe} - \omega_{ll}$, and the difference between the second-excited and first-excited states, denoted as $\mathrm{\Delta_{g2}} = \omega_{se} - \omega_{fe}$. A value of $\mathrm{\Delta_{g1}} \sim 0$ indicates the presence of a gapless mode between $\omega_{ll}$ and $\omega_{fe}$, while $\mathrm{\Delta_{g2}} \sim 0$ signifies a gapless mode between $\omega_{fe}$ and $\omega_{se}$. 

In this study, we fix the density-density interaction at $c_0 = 5$ and the Rabi coupling at $\Omega = 1$, while varying the spin-orbit (SO) coupling strength as $k_L = 0.5$ (regime I), $2.0$, and $4.5$ (regime II), as shown in Figs.~\ref{fig82}(a–c), respectively. Figures~\ref{fig82}(a i) and~\ref{fig82}(a ii) demonstrate that, when simultaneously varying the spin-exchange interaction $c_2$ and quasi-momentum $q_x$ with a fixed $c_0 = 5$, no bandgap exists ($\mathrm{\Delta_{g1}} = \mathrm{\Delta_{g2}} = 0$) only for $c_2 > 0$ (AFM interactions). This gapless state persists as $c_2$ increases, indicating no gap between $\omega_{fe} - \omega_{ll}$ and $\omega_{se} - \omega_{fe}$. Conversely, for $c_2 < 0$ (FM interactions), only gapped modes are observed between all branches.

Similarly, upon increasing the SO coupling strength to $k_L = 2.0$, the bandgaps close only for $c_2 > 0$, with the gapless region expanding in the $c_2$ plane and maintaining its gapless nature as $c_2$ increases [see Figs.~\ref{fig82}(b i) and~\ref{fig82}(b ii)]. Next, we consider a relatively large SO coupling strength of $k_L = 4.5$. In this case, the first UAC appears with $\mathrm{\Delta_{g1}} = 0$ between the low-lying and first-excited branches only for $-2 < c_2 < 0$ (FM case), consistent with findings in SOC spin-1 ferromagnetic BECs~\cite{Gangwar2024}. This first UAC persists even in the $c_2 > 0$ regime [see Fig.~\ref{fig82}(c i)]. However, a second UAC emerges for $c_2 > 0$ [see Fig.~\ref{fig82}(c ii)]. Thus, a relatively strong SO coupling exhibits two gapless UACs when $c_2 > 0$, whereas a single gapless UAC occurs for $-2 < c_2 < 0$.

In conclusion, in regime II, ferromagnetic SOC BECs exhibit a single UAC, while AFM interactions result in double UACs between the branches. In contrast, regime I remains stable, displaying two gapless stable avoided crossings for $c_2 > 0$ and only gapped modes for $c_2 < 0$.
\section{Numerical Simulation}
\label{sec:8}
In this section, we present the numerical simulation results to understand the nature of the dynamical stability of the condensate. We obtain the ground state of the condensate using the imaginary-time-propagation (ITP) method and then evolve it using the real-time-propagation method (RTP) by quenching the trap strength. We use the split-step-Crank-Nicolson scheme to implement both ITP and RTP methods~\cite{Muruganandam2009,Ravisankar2021}. We consider space grid $[-32, 32]$ with the space step $dx = 0.05$ in both ITP and RTP. The considered time step is $dt = 0.00025$ and $dt = 0.0005$ for ITP and RTP, respectively. Here, we choose two quantum phases in the two different SOC-BECs, namely, FM and AFM BECs. %

\subsection{Dynamics of ferromagnetic SOC-BECs}
In this section we present the dynamics of the ground state for the two regimes of the ferromagnetic SOC-BECs.

\begin{figure}
\begin{centering}
\centering\includegraphics[width=0.99\linewidth]{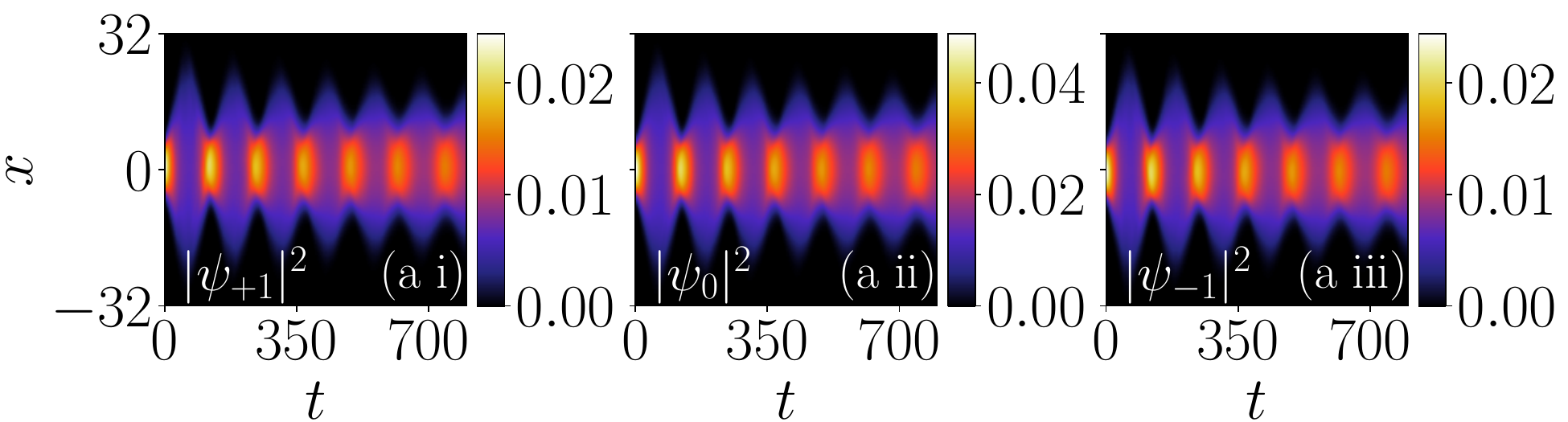}
\centering\includegraphics[width=0.99\linewidth]{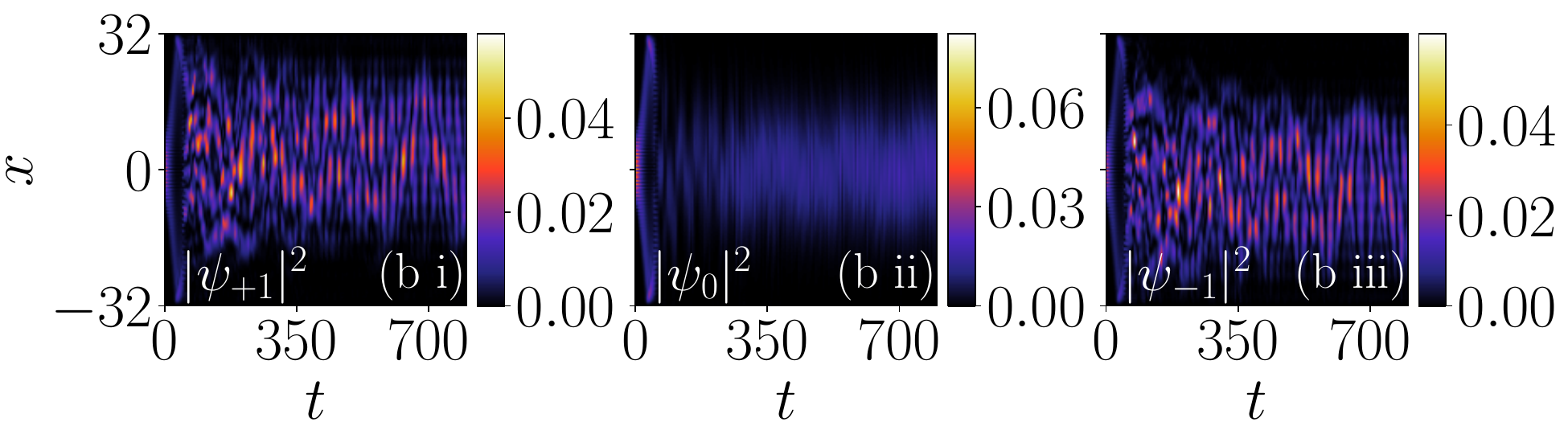}
\caption{{Time evolution of the condensate in $x- t$ plane for $\vert\psi_{+1}\vert^{2}$, $\vert \psi_{0}\vert^{2}$, $\vert\psi_{-1}\vert^{2}$ components of the condensate for (top row) $(k_{L}, \Omega)$ = (0.5, 1.0), and  (bottom row) $(k_{L}, \Omega)$ = (4.5, 1.0), upon quenching the trap strength to one-third of its initial value. The interaction strengths are $c_{0} = 5.0$ and $c_{2} = -2.0$. In the top row, the density profile shows stable behaviour throughout, showing the dynamical stability of the condensate. In the bottom row, the density profile holds its SW nature for a while; at a later time, $\vert\psi_{\pm 1}\vert^{2}$ fragments in small domains and show decay in amplitude while the zeroth component diminishes at first, further gains the amplitude.}}
 \label{t-fig14}
\end{centering}
\end{figure}

For Regime I ($k_L = 0.5$, $\Omega = 1.0$), we generate the PW phase of the FM interaction ground-state phase using the parameters $c_0 = 5.0$, $c_2 = -2.0$, $k_L = 0.5$, and $\Omega = 1.0$ under a harmonic trap with strength $\lambda = 0.10$. After obtaining the ground state, we apply an instantaneous quench in the trap strength, triggering the dynamics of the condensate, which are computed using real-time propagation. In Figs.~\ref{t-fig14}(a i–a iii), we present the dynamical evolution of the density profiles for the three components of the condensate, revealing breathing oscillations. This characteristic response of the condensate to perturbation confirms the real eigenfrequency for these parameters, as illustrated in Fig.~\ref{t-fig0}(a).


For regime II ($k_{L} = 4.5, \Omega = 1.0$), we obtain the ground state with the interaction strength $c_{0} = 5.0$, $c_{2} = -2.0$, and considering SO and Rabi couplings strength as $k_{L} = 4.5$ and $\Omega = 1.0$, respectively, which is a stripe wave (SW) phase. Further, in time evolution, perturbing the trap strength of the density profile of the condensate changes its shape and amplitude. Here, we discuss it in the $x-t$ plane. The density profile evolves as the SW in the beginning, but $t>0$, densities $\lvert \psi_{\pm 1} \rvert^2$ fragments into several small domains, while the zeroth component density diminishes up to $t\approx 400$, and afterwards $\lvert \psi_0\rvert^2$ has very week growth, further $t>400$, all the components exhibit an immiscible like nonlinear wave patterns. Moreover, we find that in the absence of magnetization ($m=0$), the condensates remain polarized in the $\pm x$ directions, which is not observed in regime I [see Figs.~\ref{t-fig14}(b i - b iii)]~\cite{Mithun2019,Gangwar2024}, due to the appearance of instability as shown in  Fig.~\ref{t-fig0}(b). Overall, we find that the numerical simulation confirms the dynamical instability of the condensate, which is in line with the collective excitation spectrum calculation.   

\subsection{Dynamics of anti-ferromagnetic SOC-BECs}
\label{afcasei}
\begin{figure}
\begin{centering}
\centering\includegraphics[width=0.99\linewidth]{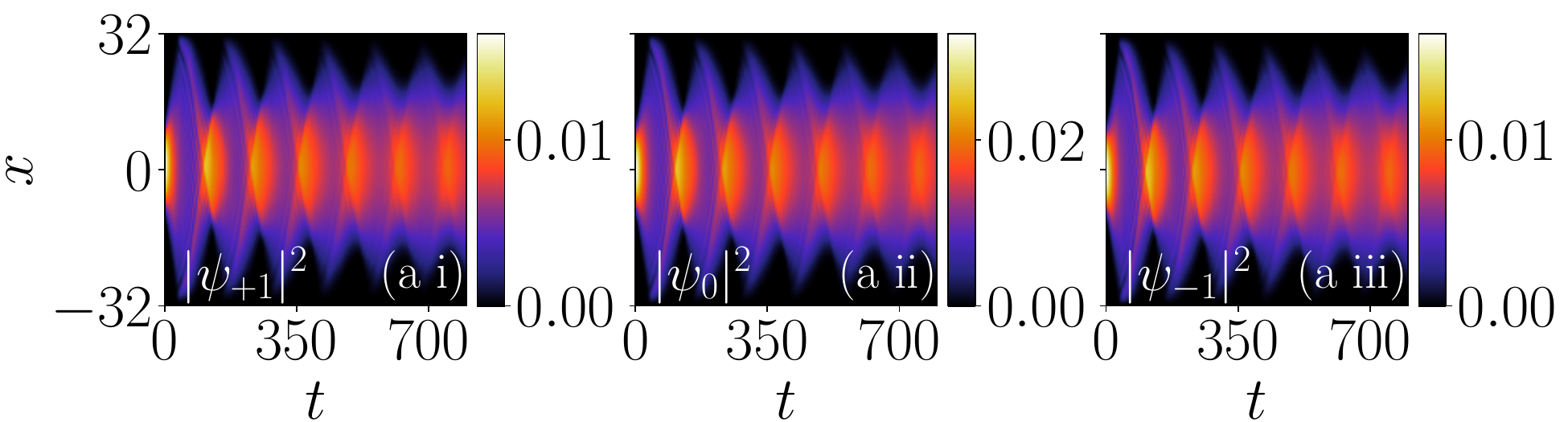}
\centering\includegraphics[width=0.99\linewidth]{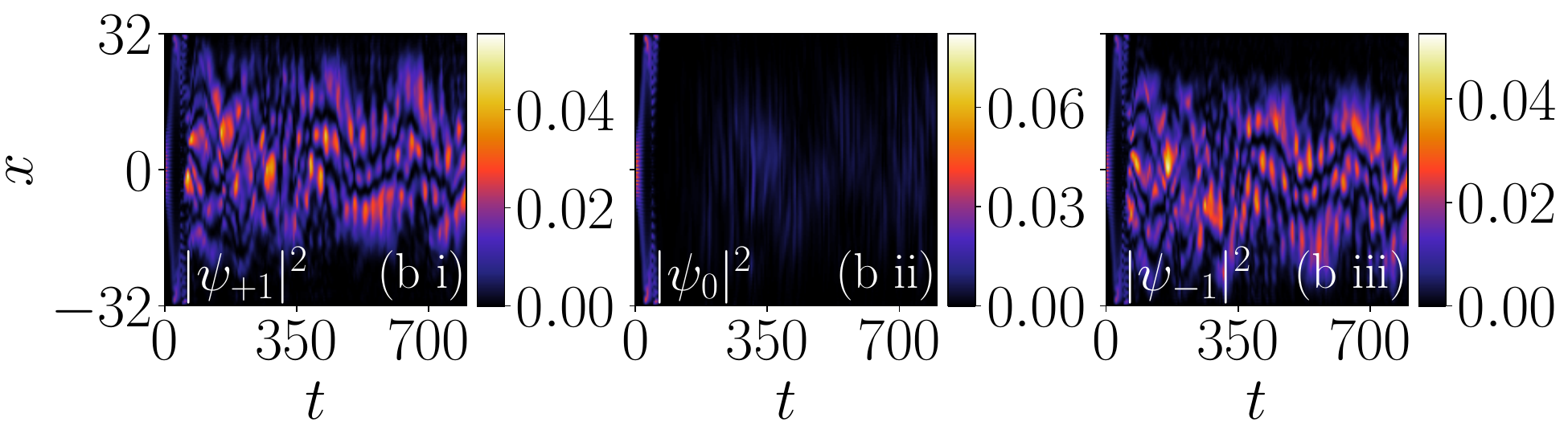}
\caption{{Dynamical evolution of the condensate in $x- t$ plane for $\vert\psi_{+1}\vert^{2}$, $\vert \psi_{0}\vert^{2}$, $\vert\psi_{-1}\vert^{2}$ components of the condensate for (top row) $k_{L} = 0.5, \Omega = 1.0$ and (bottom row) $k_{L} = 4.5, \Omega = 1.0$, by quenching the trap strength to one-third of its initial value. The interaction strengths are $c_{0} = 5.0$, and $c_{2} = 5.0$. In the PW phase, the density of the condensate shows stable breather-like dynamics. The SW phase holds its shape in the beginning, further $\vert\psi_{\pm 1}\vert^{2}$ components fragments in several small domains, and zeroth components start diminishing, confirming the dynamical instability of the condensate.}}
 \label{t-fig13}
\end{centering}
\end{figure}

First, we consider the dynamics of the  Regime I ($k_{L} = 0.5, \Omega = 1.0$) also yields the PW phase of the AFM interaction of the SOC-BECs. Initially, we generate the ground state with equal interaction strengths $c_{0} = 5.0$, $c_{2} = 5.0$, keeping the Rabi and SO couplings as $\Omega = 1.0$ and $k_{L} = 0.5$ respectively. We then generate the dynamics of the condensate by quenching the trap strength. The dynamical evolution of the condensate is obtained by using the RTP of the GP equation. In Figs.~\ref{t-fig13}(a i - a iii) we show the densities of the three component, $\lvert \psi_{+1}\rvert^2$, $\lvert \psi_0\rvert^2$, and $\lvert \psi_{-1}\rvert^2$, respectively. During temporal evolution, the density shows stable breather oscillation. This particular feature of the condensate complements the real nature of the collective excitation spectrum corresponding to this regime as presented in Fig.~\ref{t-fig00}(a). This constant behaviour of the density profile during dynamical evolution and stable energy complements the dynamically and energetically stable nature of the condensate.

In Figs.~\ref{t-fig13}(b i - b iii) we show the dynamical evolution of the density component for the regime (ii) ($k_{L} = 4.5,  \Omega = 1.0$) attained after the quench of potential strength for the ground state prepared with interaction strengths $c_{0} = 5.0$, $c_{2} = 5.0$. We find that the density profile, which has a stripe wave nature for all the components at $t=0$, starts getting fragmented into several small domains upon evolution. While $\pm 1$ components get polarized [see Figs.~\ref{t-fig13}(b i),(b iii)], the zeroth component density starts diminishing [see Fig.~\ref{t-fig13}(b ii)]~\cite{Matus2008,Kronjager2010,Tasgal2015}. This particular dynamical feature of the condensate complements the presence of the double UAC, which is present with both low-lying and second-excited branches, resulting in $I_{o}$ type of instabilities and thus making the condensate dynamically unstable. It is worth noting that a similar domain formation has been realized in the presence of a weak Zeeman coupling~\cite{Kronjager2010}. 


\section{Summary and Conclusions}
\label{sec:9}
We have studied the stability of various phases in spin-orbit coupled spin-1 SOC-BECs with FM and AFM interactions. The Bogoliubov-de Gennes theory was employed to compute the eigenspectrum of the condensate.

For ferromagnetic interactions in the low spin-orbit (SO) coupling regime ($k_{L}^{2} < \Omega$), the eigenspectrum exhibits real eigenfrequencies with a gap between the branches, showing phonon modes in the low-lying branch. The eigenvectors corresponding to the low-lying branch components approach the same value at $q_{x} \approx 0$, confirming the presence of phonon modes. In the high SOC regime ($k_{L}^{2} > \Omega$), we observe multi-band imaginary eigenfrequencies in the low-lying branch. For FM interactions, we find the emergence of an unavoided crossing (UAC) between the low-lying and first-excited branches, indicating the presence of $I_{o}$-type instability in each branch. The eigenvectors in this regime exhibit spin-density-like mixed modes arising from the multi-band eigenfrequencies in quasi-momentum space. At the wave number regime where UAC occurs, the eigenvector shows out-of-phase behaviour for all the components of the condensate.

For AFM interactions with low SOC, we observe stable avoided crossings between the low-lying and first-excited branches, as well as between the first- and second-excited branches. At these stable avoided crossings, the eigenvectors of all components display flipping characteristics. In the high-SOC regime ($\Omega < k_L^2$), we identify multi-band imaginary eigenfrequencies in the low-lying and first-excited branches, accompanied by single-band instability in the second-excited branch. Evidence of $I_o$-type instability is observed in the excitation spectrum, primarily originating from the first UAC between the low-lying and first-excited branches and the second UAC between the first- and second-excited branches. Notably, for AFM interactions, the first-excited branch exhibits a double UAC, a feature not present in the ferromagnetic case. The eigenvectors for AFM interactions reveal spin-density-like mixed modes in the low-lying and first-excited branches, while the second-excited branch transitions from density to spin modes due to single-band instability. The presence of UAC and double UAC induces out-of-phase behaviour in the respective components.

Further investigation into instability in momentum space reveals that while the instability region is symmetric in the SOC range, it shows an asymmetric character in the Rabi coupling plane. For fixed values of Rabi coupling (positive or negative), both FM and AFM SOC-BECs are destabilized only for high SOC ($k_{L}^{2} > \Omega$). In the case of weak SOC in AFM BECs, instability and UAC are observed within the range $\Omega < 3$, while the system remains stable otherwise. In contrast, the FM condensate remains stable in this regime. Both systems exhibit double UAC when $\Omega \lesssim 0$. For larger SOC, AFM interactions show UAC for $\Omega \gtrsim 6$, with double UAC appearing for $\Omega \lesssim 3$. The ferromagnetic condensate, however, remains unstable throughout the entire Rabi coupling range, with UAC observed for $\Omega \lesssim 1$ and double UAC for $\Omega \lesssim -2$. In the $k_L-\Omega$ phase plane, both systems exhibit stability only in regime I; instability prevails in all other regimes, driven by the SO coupling strength and multi-band instability.  Finally, we have investigated the variation of the band gaps between low-lying  excitation branches in the interaction parameter space, noting that AFM SOC-BECs consistently showed double UAC, while ferromagnetic interactions displayed single UAC in the eigenspectrum.

We have complemented the dynamical stability of the condensate obtained using the BdG analysis through the mean-field GPEs for both the low (regime I) and high (regime II) SOC regimes. For low SOC, we have reported the stable breather pattern of the condensate with FM and AFM interactions, which very well complement the dynamically stable nature of the condensate as shown using the BdG analysis. At high SOC, for which the ground state is of  striped wave phase nature exhibits dynamically fragmented condensate upon evolution for both the interactions. This fragmentation is more pronounced with in the AFM interaction which may be attributed to the presence of instabilities across all branches. In AFM SOC-BECs, the fragmentation is accompanied with the complex domain formation in which the amplitude of the component remains unchanged, however shape of the condensate fluctuates. In contrast, for FM SOC-BECs shows decay in the density profile, evidenced by the reduced amplitude and immiscibility of the condensates.

It would be intriguing to extend the current work to the higher dimensions (2D and 3D), where SOC-BECs can exhibit a diverse range of  interesting topological and supersolid states~\cite{Li2017,Putra2020}. Controlling the gap between the low lying excited state presented in the paper could be valuable for developing  novel quantum material and technology using the ultracold atomic systems. 

\acknowledgments
We gratefully acknowledge our Param-Ishan supercomputing facility (IITG), where all numerical simulations were performed. S.K.G gratefully acknowledges a research fellowship from MoE, Government of India. R.R. acknowledges the postdoctoral fellowship supported by Zhejiang Normal University, China, under Grant No. YS304023964. The work of P.M. is supported by MoE RUSA 2.0 (Bharathidasan University - Physical Sciences).

\onecolumngrid
\appendix
\section{Relevant terms of the BdG matrix of collective excitations}
\label{matrx:BdG}

In this appendix, we provide an explicit form of the matrix elements of the BdG matrix equation~(\ref{bdgmatrix}). The matrix elements of Eq.~(\ref{bdgmatrix}) read as

\begin{align}
H_{+}  &= \frac{q_{x}^{2}}{2} + c_{0}(2\phi_{+1}^{2} + \phi_{0}^{2} +\phi_{-1}^{2}) + c_{2}(2\phi_{+1}^{2} + \phi_{0}^{2} -\phi_{-1}^{2}) \\ H_{0} &=  \frac{q_{x}^{2}}{2} + c_{0}(\phi_{+1}^{2} + 2\phi_{0}^{2} +\phi_{-1}^{2}) + c_{2}(\phi_{+1}^{2} +\phi_{-1}^{2})
\\
 H_{-} &= \frac{q_{x}^{2}}{2} + c_{0}(\phi_{+1}^{2} + \phi_{0}^{2} +2\phi_{-1}^{2}) + c_{2}(2\phi_{-1}^{2} + \phi_{0}^{2} -\phi_{+1}^{2})
\\
 \mu_{+}\phi_{+1} &= c_{0}(\phi_{+1}^{2} + \phi_{0}^{2} +\phi_{-1}^{2}) \phi_{+1} + c_{2}(\phi_{+1}^{2} + \phi_{0}^{2} -\phi_{-1}^{2}) \phi_{+1} + c_{2} \phi_{0}^{2} \phi_{-1}^{*} +  \frac{\Omega}{\sqrt{2}} \phi_{0}
\\
 \mu_{0} \phi_{0} &= c_{0}(\phi_{+1}^{2} + \phi_{0}^{2} +\phi_{-1}^{2}) \phi_{0} + c_{2}(\phi_{+1}^{2} + \phi_{-1}^{2})\phi_{0} + 2 c_{2}\phi_{0}^{*}\phi_{+1}\phi_{-1} +  \frac{\Omega}{\sqrt{2}}(\phi_{+1}+\phi_{-1}) 
\\
 \mu_{-} \phi_{-1} &= c_{0}(\phi_{+1}^{2} + \phi_{0}^{2} +\phi_{-1}^{2}) \phi_{-1} + c_{2}(\phi_{-1}^{2} + \phi_{0}^{2} -\phi_{+1}^{2}) \phi_{-1} + c_{2} \phi_{0}^{2} \phi_{+1}^{*} +  \frac{\Omega}{\sqrt{2}} \phi_{0}
\end{align}

\begin{align}
  \mathcal{L}_{12}  &= C^{+}\phi_{+1}^{2}; \;\; \mathcal{L}_{13}= C^{+}\phi_{0}^{*}\phi_{+1}- \frac{k_{L}}{\sqrt{2}}i q_{x} + 2c_{2}\phi_{0}\phi_{-1}^{*}+ \frac{\Omega}{\sqrt{2}}; \;\; \mathcal{L}_{14}=C^{+}\phi_{0}\phi_{+1}; \nonumber \\ 
 \mathcal{L}_{15} &=C^{-}\phi_{-1}^{*}\phi_{+1} 
  \mathcal{L}_{16}  =C^{-}\phi_{-1}\phi_{+1} +c_{2} \phi_{0}^{2}; \;\; \mathcal{L}_{21}=-C^{+}\phi_{+1}^{*2}; \;\; \mathcal{L}_{23}=-C^{+}\phi_{0}^{*}\phi_{+1}^{*};   \nonumber  \\
 \mathcal{L}_{24} &= -C^{+}\phi_{0}\phi_{+1}^{*} + \frac{k_{L}}{\sqrt{2}}i q_{x} -2c_{2}\phi_{0}^{*}\phi_{-1}-\frac{\Omega}{\sqrt{2}};
  \mathcal{L}_{25}  =-C^{-}\phi_{-1}^{*}\phi_{+1}^{*} - c_{2}\phi_{0}^{*2}; 
  \nonumber  \\ 
 \mathcal{L}_{26} &=-C^{-}\phi_{-1}\phi_{+1}^{*};  \mathcal{L}_{31} = C^{+}\phi_{+1}^{*}\phi_{0} + 2c_{2}\phi_{0}^{*}\phi_{-1} +\frac{k_{L}}{\sqrt{2}}i q_{x} +\frac{\Omega}{\sqrt{2}}; \;\; \mathcal{L}_{32}=C^{+}\phi_{+1}\phi_{0}; \nonumber  \\ 
 \mathcal{L}_{34} &=c_{0}\phi_{0}^{2} +2 c_{2}\phi_{+1}\phi_{-1}; \;\;  \mathcal{L}_{35} = C^{+}\phi_{-1}^{*}\phi_{0} + 2c_{2}\phi_{0}^{*}\phi_{+1} -\frac{k_{L}}{\sqrt{2}}i q_{x} +\frac{\Omega}{\sqrt{2}}; \;\; \mathcal{L}_{36}=C^{+}\phi_{-1}\phi_{0}; \nonumber  \\ 
 \mathcal{L}_{41} &=-C^{+}\phi_{+1}^{*}\phi_{0}^{*} 
  \mathcal{L}_{42}  = -C^{+}\phi_{+1}\phi_{0}^{*}-\frac{k_{L}}{\sqrt{2}}i q_{x}- 2c_{2}\phi_{0}\phi_{-1}^{*}-\frac{\Omega}{\sqrt{2}}; \;\; \mathcal{L}_{43}=-c_{0}\phi_{0}^{*2}-2 c_{2}\phi_{+1}^{*}\phi_{-1}^{*}; \nonumber  \\
 \mathcal{L}_{45} &=-C^{+}\phi_{-1}^{*}\phi_{0}^{*}; \;\; 
  \mathcal{L}_{46} = -C^{+}\phi_{-1}\phi_{0}^{*} + \frac{k_{L}}{\sqrt{2}} i q_{x}- 2c_{2}\phi_{0}\phi_{+1}^{*}- \frac{\Omega}{\sqrt{2}}; \;\; \mathcal{L}_{51}=C^{-}\phi_{+1}^{*}\phi_{-1}; \nonumber  \\ 
 \mathcal{L}_{52} &=C^{-}\phi_{+1}\phi_{-1}+c_{2}\phi_{0}^{2}; \;\;  
  \mathcal{L}_{53} = C^{+}\phi_{0}^{*}\phi_{-1}+ \frac{k_{L}}{\sqrt{2}}i q_{x} + 2c_{2}\phi_{0}\phi_{+1}^{*} + \frac{\Omega}{\sqrt{2}}; \;\; \mathcal{L}_{54}=C^{+}\phi_{0}\phi_{-1}; \nonumber  \\ 
 \mathcal{L}_{56} &=C^{+}\phi_{-1}^{2}; \;\; \mathcal{L}_{61}= -C^{-}\phi_{+1}^{*}\phi_{-1}^{*}-c_{2}\phi_{0}^{*2}; 
  \mathcal{L}_{62} =- C^{-}\phi_{+1}\phi_{-1}^{*}; \;\; \mathcal{L}_{63}=-C^{+}\phi_{0}^{*}\phi_{-1}^{*}; \nonumber  \\ 
 \mathcal{L}_{64} &= -C^{+}\phi_{0}\phi_{-1}^{*}-\frac{k_{L}}{\sqrt{2}}i q_{x}-2c_{2}\phi_{0}^{*}\phi_{+1}-\frac{\Omega}{\sqrt{2}}; \;\; \mathcal{L}_{65}= -C^{+} \phi_{-1}^{*2}. \nonumber 
\end{align}
Also,
\begin{equation}
C^{+}\equiv c_{0} + c_{2}, \;C^{-}\equiv c_{0} - c_{2}  \nonumber 
\end{equation}
The coefficients for the BdG characteristic equation~(\ref{bdgex}) are given as follows:
\begin{align}
 b &= -5 \Omega^{2}-4 c_{2}^{2}-(2 k_{L}^{2}+3 \Omega + c_{0})q_{x}^{2}-\frac{3}{4}q_{x}^{4}+ c_{2}(8 \Omega + q_{x}^{2}), \\
 c &= 4 \Omega^{4}+ \Omega \left[(2 \Omega ( k_{L}^{2} + 3 \Omega ) - ( k_{L}^{2} -5 \Omega ) c_{0}\right]q_{x}^{2} + 4 c_{2}^{3} q_{x}^{2}  + \frac{1}{2}\left[2 k_{L}^{4} + 9 \Omega^{2} + ( k_{L}^{2} + 6 \Omega )c_{0} \right] q_{x}^{4} \nonumber  \\ & + \frac{1}{2}(3 \Omega + c_{0}) q_{x}^{6}  + \frac{3}{16} q_{x}^{8} + 4 c_{2}^{2}\left[\Omega^{2}+ ( k_{L}^{2}-\Omega + c_{0} ) q_{x}^{2} \right] \nonumber  \\ & - \frac{1}{2} c_{2} \left[16 \Omega^{3} +2 \Omega ( 7 k_{L}^{2} + 5 \Omega + 8 c_{0} )q_{x}^{2} +( k_{L}^{2} + 6 \Omega +4 c_{0} )q_{x}^{4}+ q_{x}^{6} \right], \\
 d &= -\frac{1}{64} q_{x}^{2}\left[ ( 4 k_{L}^{2} - 4 \Omega - q_{x}^{2} )( 2 \Omega + q_{x}^{2} ) + c_{2}( -8 k_{L}^{2} + 8 \Omega + 4 q_{x}^{2} )\right]  \bigg[-4 c_{0}\bigg(8 \Omega^{2} -2 (k_{L}^{2}-3 \Omega)q_{x}^{2} \nonumber  \\ 
& + q_{x}^{4} - 4 c_{2}(2 \Omega + q_{x}^{2})\bigg) +(2 \Omega + q_{x}^{2})\bigg(-16 \Omega c_{2} + 16 c_{2}^{2}-q_{x}^{2}(-4 k_{L}^{2} + 4 \Omega + q_{x}^{2})\bigg)\bigg].
\end{align}

\twocolumngrid
\bibliography{ref.bib}

\end{document}